%% file: main_IM.tex
\documentclass[aps,prx,twocolumn,superscriptaddress,10]{revtex4-2}
\usepackage{cmbright}
\DeclareFontShape{OT1}{cmss}{m}{it}{<->ssub*cmss/m/sl}{}

\usepackage{amsmath}
\usepackage{amssymb}
\DeclareUnicodeCharacter{2212}{\textendash}
\usepackage{graphicx}% Include figure files
\usepackage{dcolumn}% Align table columns on decimal point
\usepackage{bm}% bold math
\usepackage[version=4]{mhchem} % chemical formulas and elements
\usepackage{graphicx}
\usepackage{amsmath}
\usepackage{nicematrix}
\usepackage{textcomp, gensymb}
\usepackage[bb=boondox]{mathalfa}
\usepackage{subfiles}
\usepackage{tabularx} %fancy tables with fixed width
\usepackage{overpic}
\usepackage[linktocpage=true,
  colorlinks=true, 
  pdfborder={0 0 0}, 
  linkcolor=blue,
  citecolor=red,
  filecolor=yellow,
  urlcolor=blue,
  bookmarks,
  pdfauthor={},
]{hyperref}
\usepackage{blkarray}
\usepackage{lmodern}
\usepackage{comment}
\usepackage{wasysym}

\usepackage[noindentafter]{titlesec}

\titleformat{\section}
  {\normalfont\large\bfseries\MakeUppercase}{\MakeUppercase{\thesection}}{0pt}{}
\titleformat{\subsection}
  {\normalfont\bfseries}{\thesection}{0pt}{}
\titlespacing{\section}{0cm}{0.7cm}{0.01cm}
\titlespacing{\subsection}{0cm}{0.45cm}{0cm}

\usepackage{xspace}
\usepackage{tabularx}
\usepackage{array}   
\newcolumntype{L}{>{$}l<{$}} 
\newcolumntype{R}{>{$}r<{$}} 
\newcolumntype{C}{>{$}c<{$}} 

\usepackage[capitalise]{cleveref}
\usepackage{placeins}
\newcommand{\uvec}[1]{\hat{\mathbf{#1}}}
\newcommand\shyph{\text{-}}

\newcommand{\mzv}{0\phantom{.0}}

\begin{document}
\title{Microscopic origin of the magnetic interactions and their experimental signatures in  altermagnetic \ce{La2O3Mn2Se2}}

\author{Laura Garcia-Gassull}
\email{gassull@itp.uni-frankfurt.de}
\affiliation{Institut f\"ur Theoretische Physik, Goethe-Universit\"at Frankfurt, 60438 Frankfurt am Main, Germany}

\author{Aleksandar Razpopov}
\email{razpopov@itp.uni-frankfurt.de}
\affiliation{Institut f\"ur Theoretische Physik, Goethe-Universit\"at Frankfurt, 60438 Frankfurt am Main, Germany}

\author{P. Peter Stavropoulos}
\email{panagiotis@itp.uni-frankfurt.de}
\affiliation{Institut f\"ur Theoretische Physik, Goethe-Universit\"at Frankfurt, 60438 Frankfurt am Main, Germany}

\author{Igor I. Mazin}
\email{imazin2@gmu.edu}
\affiliation{George Mason University, Department of Physics \& Astronomy and  Quantum Science and Engineering Center,  Fairfax, USA}

\author{Roser Valent\'i}
\email{valenti@itp.uni-frankfurt.de}
\affiliation{Institut f\"ur Theoretische Physik, Goethe-Universit\"at Frankfurt, 60438 Frankfurt am Main, Germany}

\begin{abstract}

%Altermagnets (AM) are a recently introduced new type of magnets characterized by no net magnetization like antiferromagnets, but displaying a non-relativistic Zeeman splitting in reciprocal space like ferromagnets.
%One of the lately discussed models to realize AM is the inverse Lieb lattice (ILL). While it was initially suggested as  a purely theoretical construct, it occurs in real materials such as \ce{La2O3Mn2Se2}. However, AM on the ILL  requires the nearest-neighbor magnetic exchange, a 90$^\circ$ superexchange, to be {\it antiferromagnetic}, and dominant over the 180$^\circ$ next-nearest-neighbor superexchange, in apparent contradiction to the Goodenough-Kanamori-Anderson (GKA) rules. Yet, AM ordering was found to be the ground state in \ce{La2O3Mn2Se2}.
%To solve this puzzle, we perform combined \textit{ab initio} and analytical calculations and determine the reason for a large antiferromagnetic nearest-neighbor exchange, which emerges through a concerted action of direct exchange and superexchange. The seeming contradiction with the GKA rules is traced down to the multiorbital character of Mn$^{+2}$ ions. 
%Further, by calculating  magnon bands, we identify and clarify universal signatures of the exchange interactions  and suggest how they can be identified and constrained experimentally. 

Altermagnets (AM) are a recently introduced type of magnets, with no net magnetization like antiferromagnets, but displaying a non-relativistic Zeeman splitting in reciprocal space like ferromagnets.
One of the lately discussed models to realize AM is the inverse Lieb lattice (ILL). Initially suggested as  a purely theoretical construct,  the ILL occurs in real materials such as \ce{La2O3Mn2Se2}. 
However, AM on the ILL  requires 90$^\circ$ nearest-neighbor superexchange to be {\it antiferromagnetic} and dominant over the 180$^\circ$ next-nearest-neighbor superexchange, in apparent contradiction to the Goodenough-Kanamori-Anderson (GKA) rules.
Yet, AM ordering was found to be the ground state in \ce{La2O3Mn2Se2}.
Combining ab initio and analytical methods, we determine how direct exchange and superexchange act together to produce a large antiferromagnetic nearest‐neighbor coupling.
The seeming contradiction with the GKA rules is traced back to the multiorbital character of Mn$^{+2}$ ions. 
By calculating magnon bands, we identify universal signatures of the exchange interactions, suggesting experimental fingerprints.

\end{abstract}
\date{\today}
\maketitle

\section{Introduction}

Magnetism plays a pivotal role in modern technology, influencing disciplines such as data storage and the development of quantum computing. Although conventional magnetic classifications so far have dominated the scientific scene, a theoretical work describing magnetic states by utilizing spin-space symmetries \cite{LiborPRX2022} has brought a new topic to the forefront, altermagnetism (AM). 

While a number of AM candidates have been identified, most of them are g-wave AM, less beneficial for applications than d-wave ones. 
One of the recently discussed models to realize AM is the Lieb lattice 
introduced by E. Lieb in 1989~\cite{Lieb} to model the perovskite CuO$_2$ plane in high-Tc cuprates.
Originally, Lieb considered three lattice sites, corresponding to Cu (A site) and two oxygens (B sites). In the context of AM,  the {\it inverse} Lieb lattice (ILL) where B is a magnetic ion and A is a ligand (i.e., antiperovskite plane) has been proposed~\cite{Sudbo_model,Antonenko} as minimal microscopic model for altermagnetism. It has the advantage of not only being conceptually simple, but also it is d-wave, and therefore also interesting from the application point of view.  \ce{La2O3Mn2Se2} was recently brought into the limelight \cite{Fernandes} as an example of an ILL. Other materials with the same crystallographic motif, albeit less studied, have also been reported including, \ce{KV2Se2O} \cite{KVSeO_AM}, \ce{CsV2Se_{1.5}O} \cite{CsVSeO_AM}, and \ce{Na2Fe2S2O} \cite{NaFeSO_AM}, which are metallic or are likely dopable into metallicity, and at least some of them are AM. Below, we consider the best-studied  \ce{La2O3Mn2Se2} as a prototypical example.

\begin{figure*}[ht]
        \begin{overpic}[width=1.0\textwidth,percent,grid=false,tics=2]{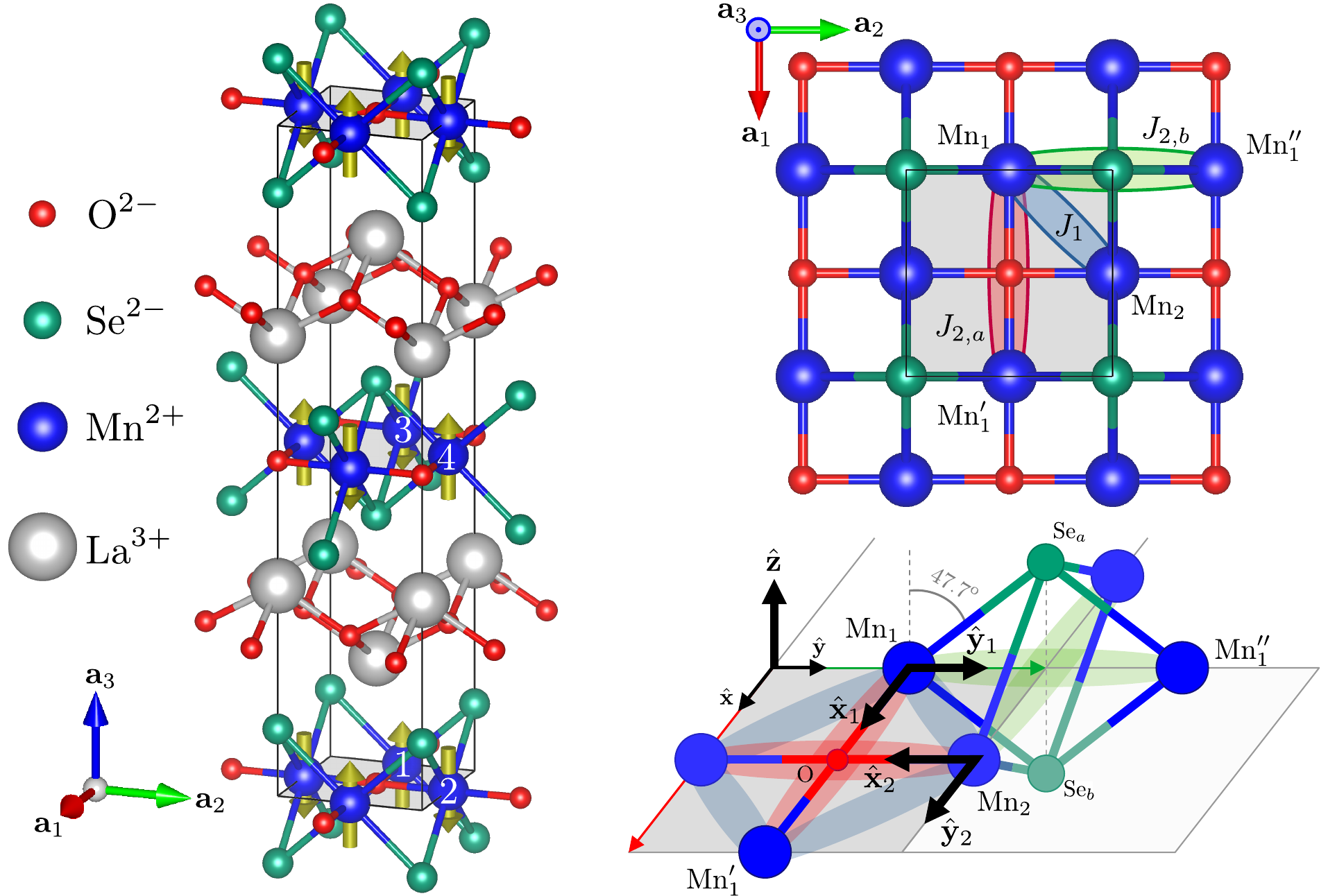}%
       \put(1,66){(a)}
       \put(46,66){(b)}
       \put(46,28){(c)}
    \end{overpic}  
    \caption{
   (a) Crystal structure of \ce{La2O3Mn2Se2} in the conventional unit cell: Mn atoms are shown in blue, O in red, La in grey and Se in green. The magnetic moment of the Mn atoms are reported to be aligned along the $\uvec{a}_3$ axis in an AFM arrangement, the so called G-AFM ordering. The \ce{Mn1} and \ce{Mn4} are connected under primitive lattice translations, as are \ce{Mn2} and \ce{Mn3}.
   (b) Top view of the [Mn$_2$OSe$_2$]$^{2-}$ layer, showing the inverse Lieb lattice formed by the Mn$_1$, Mn$_2$, and 
   %$X$=
   the ligands, O, Se. The unit cell is shown in light grey. Heisenberg exchange \(J_1\) corresponds to the NN bond between the two sub-lattices of Mn, whereas \(J_{2,a}\) and \(J_{2,b}\) correspond to same length, but not symmetry related, bonds.
    %\(B_{2,a}\) and \(B_{2,b}\) respectively. 
    (c) Oblique projection of the [Mn$_2$OSe$_2$]$^{2-}$ layer, showing the metal to metal bonds, up to lattice translations, as well as the metal to ligand paths considered in the SK analysis and the microscopic origin of the exchanges. The local coordinates on the sites Mn$_1$ and Mn$_2$ are indicated by the indices 1 and 2, while the global coordinate frame has no indices. Mn$_1^\prime$
    and Mn$_1^{\prime\prime}$ denote atoms belonging to neighboring unit cells. }\label{fig:compafig}
\end{figure*}

In the inverse Lieb lattice, as mentioned above, 
there are three distinct sub-lattices, two of them with magnetic ions and one formed by the ligand atoms (see the effective lattice in Fig.~\ref{fig:compafig} (b)). 
the presence of two magnetic sub-lattices within the unit cell allows magnetic order with $\mathbf{q=0}$, a prerequisite to AM (as opposed to a CuO$_2$ layer where ordering doubles the unit cell). The requirement for a $\mathbf{q=0}$ order is the nearest neighbor exchange being antiferromagnetic and dominant.

In order to resolve the microscopic nature of the exchange interactions in \ce{La2O3Mn2Se2} we present in this work a full \textit{ab initio} analysis.
The paper is organized as follows. We first analyze the crystal symmetry of \ce{La2O3Mn2Se2} and discuss a possible realization of altermagnetic order.
We proceed with density functional theory (DFT) calculations and 
extract the exchange parameters by mapping the energies of different spin configurations to the Heisenberg model. 
%Surprisingly, we find contradicting results to reported exchange couplings, but our results agree well with the GKA rules.
We rationalize the calculated exchange parameters by  a Slater-Koster (SK) analysis and discuss virtual paths that give rise to these exchanges. Next, we address the experimentally observed magnetic anisotropy by performing a symmetry analysis of a general anisotropic bilinear spin Hamiltonian for \ce{La2O3Mn2Se2}.
As a last step, we determine the magnetic ground state and magnon spectrum at the level of linear spin wave theory and discuss the consequences in experimental observations.

\section{Methods}\label{sec:methods}
\subsection{Crystal Structure and Computational Details}\label{sec:comp_det}

 \ce{La2O3Mn2Se2} crystallizes in the space group \(I4/mmm\) (number 139) with associated crystal point group \(D_{4h}\) and  lattice parameters \(a=b=4.1274\)~\AA, \(c=18.8057\)~\AA~\cite{Fernandes}. 
 %It consists of 16 symmorphic point group operations, plus all primitive translations.
 Fig.~\ref{fig:compafig}(a)  shows its conventional unit cell in the orthogonal basis 
 $\mathbf{a}_1$, $\mathbf{a}_2$, $\mathbf{a}_3$ where
  [Mn$_2$OSe$_2$]$^{2-}$ layers
 %, with \ce{Mn} shown in blue, \ce{O} in red, and \ce{Se} in green, 
 are separated by sheets of [La$_2$O$_2$]$^{2+}$.
 %, with \ce{La} shown in grey.
%The crystal respects the \(I4/mmm\) space group symmetry (number 139), with the associated crystal point group \(D_{4h}\). It consists of 16 symmorphic point group operations, plus all primitive translations. 
%Focusing on the on-site symmetry of Mn atoms (see Fig.~\ref{fig:compafig}(a) for sublattice labels), 
The symmetry at the Mn$_1$ site (Fig.~\ref{fig:compafig}(b)) derives from its Wyckoff position 4c, which has local $D_{2h}$ symmetry, a subgroup of the crystal point group $D_{4h}$. This includes $C_{2\uvec{a}_i}$ rotations and inversion $I$.
%, but excludes operations like $C{4\uvec{a}_3}$, which relate Mn$_1$ to Mn$_2$.
Although Mn$_1$ and Mn$_2$ are related by $C_{4\uvec{a}_3}$, not $I$, they also connect to Mn$_4$ and Mn$_3$ via a primitive lattice translation  $\mathbf{a}^{(p)}_1=-\mathbf{a}_1/2+\mathbf{a}_2/2+\mathbf{a}_3/2$. As a result, the lattice contains two magnetic sublattices, related by $C_4$ but not by inversion -- a prerequisite for AM.

Considering the bond symmetry, the nearest-neighbor (NN) bond ($J_1$ in Fig.~\ref{fig:compafig}(b)) has $C_{2v}$ symmetry, with a $C_{2(\uvec{a}1 - \uvec{a}2)}$ primary rotation axis, but no inversion symmetry. The next-nearest-neighbor (NNN) bonds split into two symmetry-inequivalent types, $\text{NNN}_a$ and $\text{NNN}_b$ ( $J_{2,a}$ and $J_{2,b}$, respectively, in Fig.~\ref{fig:compafig}(b)).
%under any crystallographic symmetry, despite being equal in length. 
This is the origin of AM seen in the bilinear spin exchange model (see below). Both NNN types preserve $D_{2h}$ symmetry, reflecting that they bond sites of the same sublattice. Full symmetry details are provided in Appendix ~\hyperref[appx:symm_anal]{S1}.

Density functional theory (DFT) calculations were performed using the full-potential local-orbital code (FPLO) version 22.00-62~\cite{fplo,opahle1999full}. We cross-checked them against the Vienna Ab initio Simulation Package (VASP) \cite{kresse_ab_1993,kresse_ab_1994,kresse_efficiency_1996,kresse_efficient_1996} and Wien2k \cite{blaha2002wien2k,blaha2020wien2k} LAPW code.
%This version of FPLO is the one used throughout all DFT calculations.
In this work, the generalized gradient approximation (GGA) with the Perdew-Burke-Ernzerhof \cite{perdew_generalized_1996,perdew_generalized_1997} exchange-correlation functional was used. For FPLO, a \textbf{k}-mesh of 12x12x12 was used, and both the density and the energy were used as convergence conditions (\(10^{-6}\) for the former and \(10^{-8}\) Ha for the latter).

Since FPLO uses a local-orbital basis, we had no need to use external wannierization for the tight binding analysis.  In the latter, the local x-axis was set along  $\mathbf{a}_1$ for all atoms, except \ce{Mn_2}, for which it was set along $-\mathbf{a}_2$, and the local z-axis was along $\mathbf{a}_3$. The energy window for the direct hoppings was set from -7.1 to 1.37 eV, and from -2.15 to 1.37 eV for the indirect hoppings.
%The crystal structure of \ce{La2O3Mn2Se2} was taken from Ref.~\cite{Fernandes} with lattice parameters \(a=b=4.1274\)~\AA~and \(c=18.8057\)~\AA~ obtained by single-crystal X-ray diffraction at 100 K. 

The isotropic Heisenberg magnetic exchange couplings were estimated via the total energy mapping analysis (TEMA)~\cite{totalenergy_analysis0,total_analysis,total_analysis1}, by a least squares fit of the calculated energies to the spin-5/2 Heisenberg Hamiltonian.
%The DFT calculations were performed using the same FPLO code version as for the DOS and the wannierization, 22.00-62~\cite{fplo,opahle1999full}, with the GGA functional~\cite{perdew_generalized_1996}.
%The strong correlation of the localized Mn $d$-orbitals were taken into account on the level of DFT+U, as implemented in FPLO in the atomic limit~\cite{Koepernik_2009_atomic_limit}.
In all calculations the Hund's coupling $J_{H}$ was set to 1 eV and we varied the on-site Coulomb $U$.
In total, 29 different magnetic configurations were considered in three different cells of  sizes 1$\times$1$\times$1, 2$\times$2$\times$1, and 3$\times$2$\times$1 in terms of the primitive unit cell on well converged \textbf{k}-mesh.
The energy convergence criterion for each magnetic configuration is set to 10$^{-8}$ Ha.

\subsection{Slater-Koster analysis}
The Slater-Koster (SK) analysis \cite{SKanalysis1954} assumes that only two-center hopping integrals are present. This approximation is equivalent to treating every bond as an isolated diatomic molecule. We define the metal to metal $d$-orbital hopping matrix $T_{\rm Mn\shyph Mn}^{(\uvec{z})}$ as well as the metal to ligand $X$ hopping matrix $T_{{\rm Mn}\shyph X}^{(\uvec{z})}$ aligned along the $\uvec{z}$-axis. For these $\uvec{z}$-axis aligned bonds we define the SK metal-to-metal $d$-orbitals hoppings $t_{dd\sigma}\!<\!0$, $t_{dd\pi}\!>\!0$, $t_{dd\delta}\!<\!0$, as well as the metal-$d$-orbital-to-ligand-$X$-$p$-orbital hoppings $t_{X,dp\sigma}\!<\!0$, $t_{X,dp\pi}\!>\!0$ (the sign varies according to the sigh  of the overlapping orbital lobes in each case). Performing an appropriate rotation, we can derive the hopping matrix $T^{(\uvec{z})}$ for each bond of interest $T$. The details of this procedure are laid out in the Appendix~\hyperref[appx:sk]{S2}. Finally, the effective metal-to-metal SK hopping $\widetilde{T}_{\rm Mn\shyph Mn}$ is calculated perturbatively by integrating out the ligand $X$ states, 
\begin{equation}\label{eq:pert_lig_met}
    \widetilde{T}_{\rm Mn\shyph Mn} = T_{\rm Mn\shyph Mn} +  \sum_{X,p}  \dfrac{T_{{\rm Mn}\shyph X}\left|X,p\right\rangle \left\langle p,X\right|T_{{\rm Mn}\shyph X}^{\dagger}}{\Delta_{X}},
\end{equation}
where $\Delta_{X}$ is the energy difference between the metal $\rm Mn$ and ligand $X$ state.

\section{Results}

%While on-site symmetry can be directly obtained from the symmetry of Wyckoff positions, analyzing bond symmetries requires more effort. 
%We focus on nearest-neighbor (NN) and next-nearest-neighbor (NNN) bonds, which split into $\text{NNN}_a$ and $\text{NNN}_b$. We determine which $D_{4h}$ operations leave these bonds invariant (up to lattice translations).

\subsection{Electronic structure}\label{sec:DFT}

Our calculated nonmagnetic electronic structure is in agreement with previous reports \cite{xu2012first,Fernandes}. The analysis of the corresponding density of states (DOS) (Appendix~\hyperref[appx:additional_dft_res]{S3}) shows overlap between Mn $3d$- and X= O, Se $p$-ligand states around -2.0 eV indicating a significant metal-ligand hybridization. This suggests that both superexchange paths Mn-O-Mn and Mn-Se-Mn (Fig.~\ref{fig:compafig} (b)) will contribute to the magnetic exchanges.

Next, we performed spin-polarized DFT calculations and find a magnetic moment of $5\ \mu_{B}$ for \ce{Mn}, consistent with the local spin moment $S=5/2$ expected  for high spin Mn$^{2+}$.
The lowest-energy spin configuration corresponds to two antiparallel sublattices,  $\mathbf{Mn}_1\upharpoonleft \! \downharpoonright \mathbf{Mn}_2$. Since the crystal space group is body-centered, the next Lieb layer must have antiparallel magnetic moments, i.e., $\mathbf{Mn}_3||\mathbf{Mn}_2$, $\mathbf{Mn}_4||\mathbf{Mn}_1$ (see Fig.~\ref{fig:compafig}(a)). This is often called G-type AFM (G-AFM) in the literature ~\cite{Khomskii_2014}.

 Including Mn $3d$ correlation effects beyond DFT in DFT+$U$ calculations, we obtain AM spin-split bands as shown in Appendix~\hyperref[appx:additional_dft_res]{S3}.
 %consistent with Ref.~\cite{Fernandes}.
 Using the G-AFM state in the DFT+$U$ calculations with a Hubbard repulsion  $U\approx 2$ eV for the Mn $3d$ orbitals, a gap of  $0.2~\text{meV}$ is obtained in agreement with the experimentally estimated activation energy (between $0.2-0.4~\text{meV}$) from conductivity experiments~\cite{Evans,xie2013magnetic,liu2011structural}.

%In the following we first extract the Heisenberg exchanges by fitting to the DFT energies, and next we reason about the results from hoppings extracted from wannierizing.

\subsection{Effective and direct $d$-$d$ hoppings}

\begin{figure}
    \centering
        \begin{overpic}[width=1.0\linewidth,percent,grid=false,tics=5]{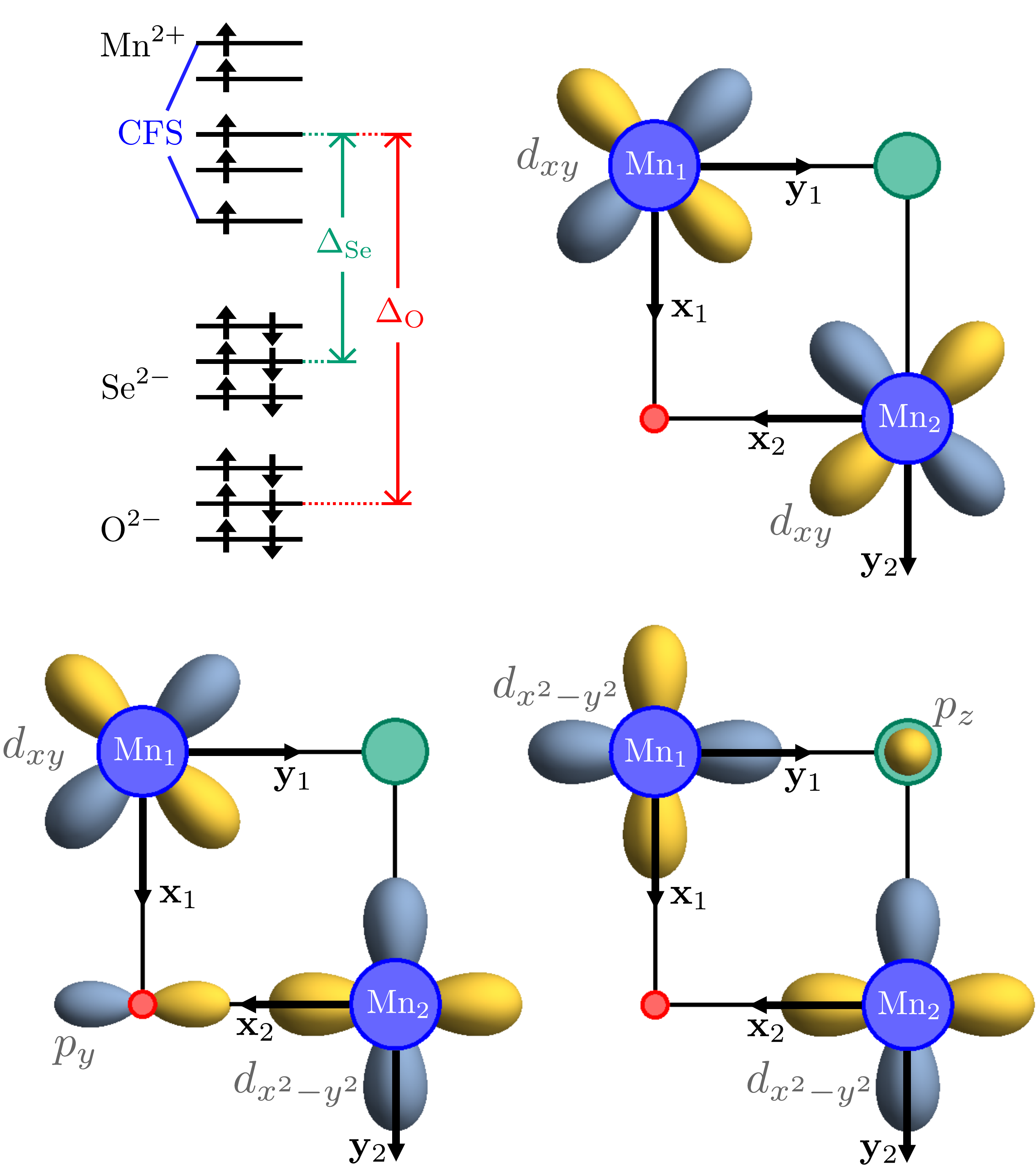}%
       \put(0,97){(a)}
       \put(42,97){(b)}
       \put(0,45){(c)}
       \put(42,45){(d)}
    \end{overpic} 
    \caption{(a) Level diagram of the metal and ligands in the ground state of \ce{La2O3Mn2Se2}, with Mn stabilizing a high spin configuration as a result of Hund's coupling $J_H$. The atomic energy difference $\Delta_X$ is shown. The crystal field splitting (CFS)  on the metal lifts the degeneracy of the $d$-orbitals completely. (b-d) Largest NN hoppings, mediated by the ligands, to form the effective hoppings $\widetilde{t}$. In panel (b) the  $\widetilde{t}_{xy\!-\!xy}$ is set by the direct overlap. The $\widetilde{t}_{(x^2-y^2)\!-\!xy}$ in panel (c) and $\widetilde{t}_{(x^2-y^2)\!-\!(x^2-y^2)}$ in panel (d) are mainly set by O and Se paths respectively.}
    \label{fig:microscopics}
\end{figure}
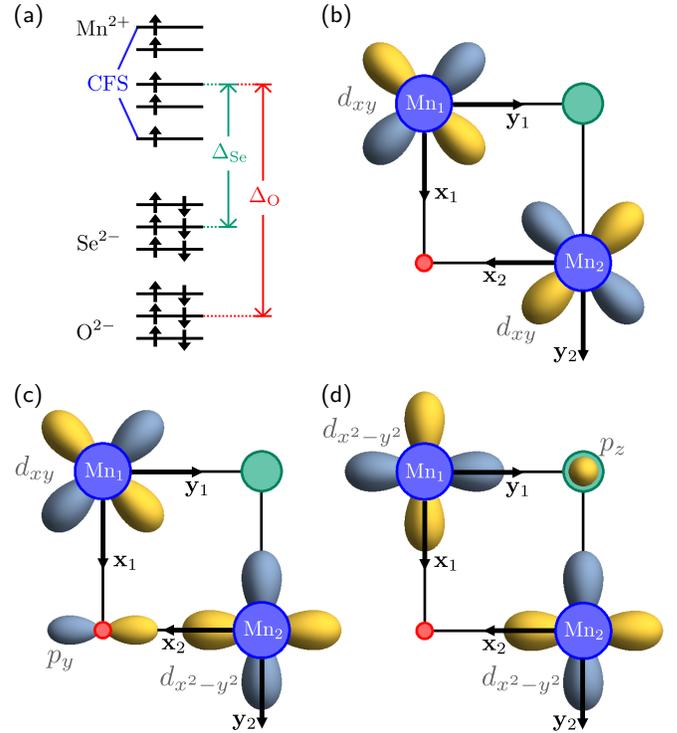

In order to unveil the origin of the sign and magnitude of the magnetic exchange parameters, we first investigate the processes involved in such an exchange, namely the ligand-assisted metal-to-metal hoppings.

To this end, we used the built-in  FPLO wannierization and extracted the hopping parameters between the atoms in the \([\text{Mn}_2\text{OSe}_2]\) layer. We consider two cases: \textit{i}) wannierization of the metal (Mn) $d$-orbitals and the ligand (O and Se) $p$-orbitals in an extended energy window, \textit{ii}) wannierization of only the metal Mn $d$-orbitals in a smaller restricted window around the Fermi level to include only the $d$-orbitals. The former case treats all orbitals on the same footing in the larger energy window resulting in bare ``direct'' hoppings between metal to metal and metal to ligand, while the latter case corresponds to ``effective'' hoppings, where the  ligand states are integrated out and effectively absorbed in the metal orbitals. Since in the latter case the resulting hoppings include direct $d-d$ hopping and the superexchange  $d-p-d$, we can determine from this study the importance of  each of these terms.
%distinguish the direct $d-d$ hopping as well as the effective correction of integrating out the ligand paths $d-p-d$.

We write the model in the natural local frame shown in Fig.~\ref{fig:compafig}(c), where the sub-lattices \ce{Mn1} and \ce{Mn2} have a different local frame $\uvec{x}_i\uvec{y}_i\uvec{z}_i$, $i=1,2$. We keep \ce{Mn1} aligned with the global frame $\uvec{x}_1\uvec{y}_1\uvec{z}_1\!=\!\uvec{x}\uvec{y}\uvec{z}$, and write \ce{Mn2} in the $45\degree$ rotated frame $\uvec{x}_2\uvec{y}_2\uvec{z}_2\!=\!(-\uvec{y})\uvec{x}\uvec{z}$. 
%The choice is such that 
Consequently, the local $\uvec{x}_i$-axis points from metal $\text{Mn}_i$ to the $\text{O}$, while the  $\uvec{z}$-axes remain aligned. The ligands $X\!=\!\text{O},\ \text{Se}_w$ ($w\!=\!a,b$) are kept in the global frame.

%we only wannierize the Mn d-orbitals and in the second one we include the p orbitals of the ligands (O, Se). 
%This results in two sets of hopping matrices, an effective (the former) and a direct one (latter case). By comparing these two hopping matrices, we obtain information on how the effective hoppings are renormalized by the superexchange paths through the ligands.
%Requesting only the $d$-orbital wannierization results in the effective \(\mathrm{Mn}-\mathrm{Mn}\) result as predicted from the converged DFT run, where the indirect contributions are accounted for in the projection numerically. On the other hand, including the p-orbitals of the ligands during the wannierization process results in the NN hopping matrix depending solely on the direct hoppings. 
%we also calculated the wannierization considering both the \ce{Mn} d-orbitals and the \ce{O} and \ce{Se} p-orbitals. Hence, we obtain the matrices for the direct metal to metal hopping,  \(J_1^{\text{dir}}\), without any renormalization coming from the superexchange. 
\subsubsection{Metal to metal NN hopping}
On the NN bond, we find the effective $d\!-\!d$ hoppings 
\begin{equation} \label{eq:t_B1_effective}
    \begin{array}{l}
        \widetilde{\mathbb{T}}_{\text{NN}} = \\[6pt]
        \begin{matrix}
            \begingroup
                \arraycolsep=2pt
                \begin{array}{rrrrr}
                    \makebox[\widthof{${x^2\!\!-\!y^2}$}][r]{${x^2\!\!-\!y^2}$}  &
                    \makebox[\widthof{$-74.2$}][r]{${z^2}$} & 
                    \makebox[\widthof{$-119.2$}][r]{${yz}$} & 
                    \makebox[\widthof{$55.7$}][r]{${zx}$} & 
                    \makebox[\widthof{$-249.3$}][r]{${xy}$}
                \end{array} 
            \endgroup &
            \begin{matrix} 
                \makebox[\widthof{${x^2\!\!-\!y^2}$}]{\phantom{0}} 
            \end{matrix} \\[4pt]
            \left(
            \begingroup
                \arraycolsep=2pt
                \begin{array}{rrrrr}
                    \makebox[\widthof{${x^2\!\!-\!y^2}$}][r]{$71.5$} 
                          & -74.2 &  \mzv  & \mzv  & -249.3 \\
                    -74.2 & -61.2 &  \mzv  & \mzv  &  -61.0 \\
                    \mzv  & \mzv  & -119.2 &  -1.0 &  \mzv  \\
                    \mzv  & \mzv  &    1.0 &  55.7 &  \mzv  \\
                    249.3 &  61.0 &  \mzv  & \mzv  &  301.1 
                \end{array}
            \endgroup
            \right) &
            \begingroup
                \arraycolsep=-20pt
                \begin{array}{l} 
                    {x^2\!\!-\!y^2} \\ {z^2} \\ {yz} \\ {zx} \\ {xy} 
                \end{array}
            \endgroup
        \end{matrix}
    \end{array}
\end{equation}
between metal sites \ce{Mn1}-\ce{Mn2} in their respective local frames, in meV.
%,with the matrix written in the orbital basis \(\{d_{x^2\!-\!y^2}\), \(d_{z^2}\), \(d_{yz}\), \(d_{zx}\), \(d_{xy}\}\).  
This matrix respects the \(C_{2v}\) point group, i.e., the self-mapping subgroup of this bond.
%The largest hopping occurs between $d_{xy}\!-\!d_{xy}$ (Fig.~\ref{fig:microscopics} (b)), which is unsurprising, since the orbitals face each other. The second most dominant hopping occurs between the $d_{x^2-y^2}\!$ and $\!d_{xy}$ orbitals (Fig.~\ref{fig:microscopics} (c)). 

There are three main %$d\!-\!p\!-\!d$ 
paths that contribute to the effective $d\!-\!d$ hopping $\widetilde{\mathbb{T}}_{\text{NN}}$: 
\textit{i}) direct \(\mathrm{Mn}_1\shyph\mathrm{Mn}_2\) with a bond length of \(\sim 3\text{ \AA}\) (Fig.~\ref{fig:microscopics} (b)), with $d_{xy}-d_{xy}$ being the largest since the orbitals face each other, 
\textit{ii}) indirect planar \(\mathrm{Mn}_1\shyph\mathrm{O}\shyph\mathrm{Mn}_2\) path in a \(90\degree\) geometry and $\mathrm{Mn}\shyph\mathrm{O}$ length of \(\sim 2\text{ \AA}\) (Fig.~\ref{fig:compafig} (c) and Fig.~\ref{fig:microscopics} (c)),  
\textit{iii}) indirect off plane \(\mathrm{Mn}_1\shyph\mathrm{Se}_{w}\shyph\mathrm{Mn}_2\) ($w\!=\!a,b$) paths with the approximately \(\sim 60 \degree\) angle and a $\mathrm{Mn}\shyph\mathrm{Se}_w$ length of \(\sim 2.8 \text{ \AA}\) (Fig.~\ref{fig:compafig} (c) and Fig.~\ref{fig:microscopics} (d)). 

We can compare the effective $d-d$ hoppings from Eq.~(\ref{eq:t_B1_effective}) with the direct $d\!-\!d$ hoppings 
\begin{equation} \label{eq:t_B1}
    \begin{array}{l}
        \mathbb{T}_{\text{NN}} = \\[6pt]
        \begin{matrix}
            \begingroup
                \arraycolsep=2pt
                \begin{array}{rrrrr}
                    \makebox[\widthof{$-125.1$}][r]{${x^2\!\!-\!y^2}$}  &
                    \makebox[\widthof{$-125.1$}][r]{${z^2}$} & 
                    \makebox[\widthof{$99.0$}][r]{${yz}$} & 
                    \makebox[\widthof{$-117.4$}][r]{${zx}$} & 
                    \makebox[\widthof{$-110.8$}][r]{${xy}$}
                \end{array} 
            \endgroup &
            \begin{matrix} 
                \makebox[\widthof{${x^2\!\!-\!y^2}$}]{\phantom{0}} 
            \end{matrix} \\[4pt]
            \left(
            \begingroup
                \arraycolsep=2pt
                \begin{array}{rrrrr}
                     -48.7 &  -125.1 &\mzv  &  \mzv  & -103.4 \\
                    -125.1 &  -58.0  &\mzv  &  \mzv  & -110.8 \\
                     \mzv  &  \mzv   & 26.0 &  -99.1 &  \mzv  \\
                     \mzv  &  \mzv   & 99.1 & -117.4 &  \mzv  \\
                     103.4 &   110.8 &\mzv  &  \mzv  &  325.2
                \end{array}
            \endgroup
            \right) &
            \begingroup
                \arraycolsep=-20pt
                \begin{array}{l} 
                    {x^2\!\!-\!y^2} \\ {z^2} \\ {yz} \\ {zx} \\ {xy} 
                \end{array}
            \endgroup
        \end{matrix}
    \end{array}
\end{equation}
in meV. 
%By comparing the $d_{xy}$-$d_{xy}$ hopping in both the direct and effective case, we can see that most of the hopping occurs through direct Mn hopping and that the paths through ligands do not renormalise it much. 
The direct $d_{xy}\!-\!d_{xy}$ hopping (Fig.~\ref{fig:microscopics}(b)) is determined by the $\sigma$-integral $t_{xy\!-\!xy}\simeq -3t_{dd\sigma}/4$, which is overall positive since a positive lobe points towards a  negative lobe in the corresponding orbitals. A Slater-Koster (SK) argument (see Appendix) shows that, to leading order, the ligands do not renormalize this direct $d_{xy}\!-\!d_{xy}$ hopping, resulting in a direct $t_{xy\!-\!xy}=325.2\text{ meV}$ and effective $\widetilde{t}_{xy\!-\!xy}=301.1~\text{meV}$ which are nearly equal. We note that the bond length of \(\sim 3\text{ \AA}\) is on the borderline of being small enough for the Mn orbitals to directly overlap, making the large direct hopping unexpected. As a reference system, Mn-based Heusler alloys \cite{wollmann2014magnetism}, where the direct Mn-Mn exchange is very important, have a \ce{Mn}-\ce{Mn} length of the order of \(2.5\text{ \AA}\).

The ligand-assisted paths come into play when considering other large effective hoppings in Eq.~\ref{eq:t_B1_effective}. The metal to ligand hopping comes at an energy cost $\Delta_X$, where $\Delta_X=\varepsilon_{\text{Mn}}-\varepsilon_{X}$ is the on-site energy difference between Mn-d and ligand $X$-p states. This can be extracted from  the on-site wannierization parameters, which we find to be $\Delta_{\text{O}} \sim 4100\text{ meV}$ and $\Delta_{\text{Se}} \sim 2784\text{ meV}$. Including the crystal field splitting (CFS) of the metal, this sets the schematic level diagram in Fig.~\ref{fig:microscopics}(a).

We examine the cross $e_g\!-\!t_{2g}$ hopping between $d_{x^2-y^2}\!-\!d_{xy}$, which is the second most dominant hopping between NN (Fig.~\ref{fig:microscopics}(c)).
By comparing the effective and the direct hopping values, we find that only $40\%$ of the effective hopping $\widetilde{t}_{(x^2-y^2)\!-\!xy}=-249.3\text{ meV}$ comes from the direct Mn-Mn hopping, $t_{(x^2-y^2)\!-\!xy}=-103.4\text{ meV}$ (compare Eq.~\ref{eq:t_B1_effective} with Eq.~\ref{eq:t_B1}). Considering the SK arguments again, the direct off-diagonal hopping $t_{(x^2-y^2)\!-\!xy}$ would be zero, which indicates that three-centre direct integrals start to become important.
Integrating out the ligands at the SK level contribute an effective correction from 
the oxygen $\sqrt{3}t_{\text{O},dp\pi}t_{\text{O},dp\sigma}/(2\Delta_{\text{O}})$, which is overall negative. The selenium paths also contribute to this hopping, but due to their off-plane geometry we find a mixed contribution $-[t_{\text{Se},dp\sigma}^2+(\sqrt{3}/2)t_{\text{Se},dp\sigma}t_{\text{Se},dp\pi}]/(2\Delta_{\text{Se}})$ with both a negative and positive term, making the overall selenium contributions small. Consequently, the effective hopping $\widetilde{t}_{(x^2-y^2)\!-\!xy}$ is mainly set by the oxygen paths. By the same arguments, the only other cross $e_g\!-\!t_{2g}$ hopping allowed, which is between $d_{z^2}\!-\!d_{xy}$, is also mainly set by the oxygen paths.

While the selenium paths mostly cancel out in the   $e_g\!-\!t_{2g}$ effective hoppings, they will dominate the corrections where the oxygen paths are negligible. This is the case for exclusively $e_g$ hoppings $\widetilde{t}_{(x^2-y^2)\!-\!(x^2-y^2)}$ (Fig.~\ref{fig:microscopics}(d)), $\widetilde{t}_{(x^2-y^2)\!-\!z^2}$, and $\widetilde{t}_{z^2\!-\!z^2}$, as well as an exclusively $t_{2g}$ hoppings $\widetilde{t}_{yz\!-\!yz}$, $\widetilde{t}_{zx\!-\!zx}$, and $\widetilde{t}_{yz\!-\!zx}$, in which cases the oxygen path with its \(90\degree\) geometry has zero effective corrections from integrating out the ligands.

\subsubsection{Metal-to-metal NNN hopping}

As mentioned above, we have two symmetry-inequivalent $\text{NNN}_a$ and $\text{NNN}_{b}$ bonds. In both cases, the metals are separated by the same distance of $\sim 4 \text{ \AA}$. The direct hopping integrals are expected to die off as $1/R^{l+l'+1}=1/R^5$ with distance \cite{heine1980tightbinding} so that $|t_{\text{NNN}}|=|t_{\text{NN}}| R_{\text{NNN}}^5/R_{\text{NNN}}^5$. Considering the largest NN hopping, we expect NNN hopping values to be at most of order $325\!\times\!3^5/4^5\sim78\text{ meV}$, about $25\%$ of the \(t_\text{NN}\) values. The largest calculated hoppings are about $30\%$ of the NN values, which is in good agreement with the expected behaviour from a distance argument. The detailed hopping matrices for $\text{NNN}_a$ and $\text{NNN}_{b}$ can be found in the Appendix \hyperref[app:eff_direct_hopp_matrices]{S3}.

Since the NNN direct hoppings are  much smaller  than  the NN ones, the effective $\text{NNN}_a$ and $\text{NNN}_{b}$ hoppings will be predominantly mediated by the ligand paths. On the $\text{NNN}_a$ bond we have the largest effective hopping between \(d_{zx}\) orbitals, mediated by the \ce{O} \(p_z\) orbital, followed closely by the hoppings between the \(d_{x^2-y^2}\) orbitals and the \(d_{xy}\) orbitals. With both  metals and ligand seating along our $x$-axis, the dominant orbitals are expected to have an $x$ component.
Similarly, on the $\text{NNN}_b$ bond, the largest effective hoppings occur between the orbitals with a $y$ component in the selected frame. The largest hopping is between \(d_{z^2}\), 
%partially 
mediated by the off-plane \(p_y\) and \(p_z\) \ce{Se} orbitals. % and partially coming from the direct Mn-Mn hopping. 
Comparing these largest effective hoppings to the direct hoppings as given in Eq.~\ref{eq:t_B1_effective} and Eq.~\ref{eq:t_B1}, we find that indeed the direct contributions are minimal.

\subsection{Isotropic magnetic exchanges}

We consider the isotropic magnetic model involving only $SU(2)$ invariant Heisenberg exchanges. Keeping up to NNN the Hamiltonian reads
\begin{equation}\label{eq:iso_spin_model}
H \!=\!\!  \displaystyle\sum\limits_{\langle i,j \rangle} J_1 \mathbf{S}_i\!\cdot\! \mathbf{S}_j \!+\!\!\! \displaystyle\sum\limits_{\substack{\gamma=a,b,\\\langle\!\langle n,n' \rangle\!\rangle_{\gamma}}} \!\!\!J_{2,\gamma}\mathbf{S}_n \!\cdot\!  \mathbf{S}_{n'},
\end{equation}
where $\langle i,j \rangle$ runs over NN bonds with $i\in\mathrm{Mn}_1$ sub-lattice and $j\in\mathrm{Mn}_2$ sub-lattice, and $\langle\langle n, n' \rangle\rangle_{\gamma}$ runs over NNN$_\gamma$ bonds with $n$ and $n'$ belonging to the same sub-lattice, with bonds indicated in Fig.~\ref{fig:compafig}(b),(c). 

\begin{figure}
    \centering
    \includegraphics[width=\linewidth]{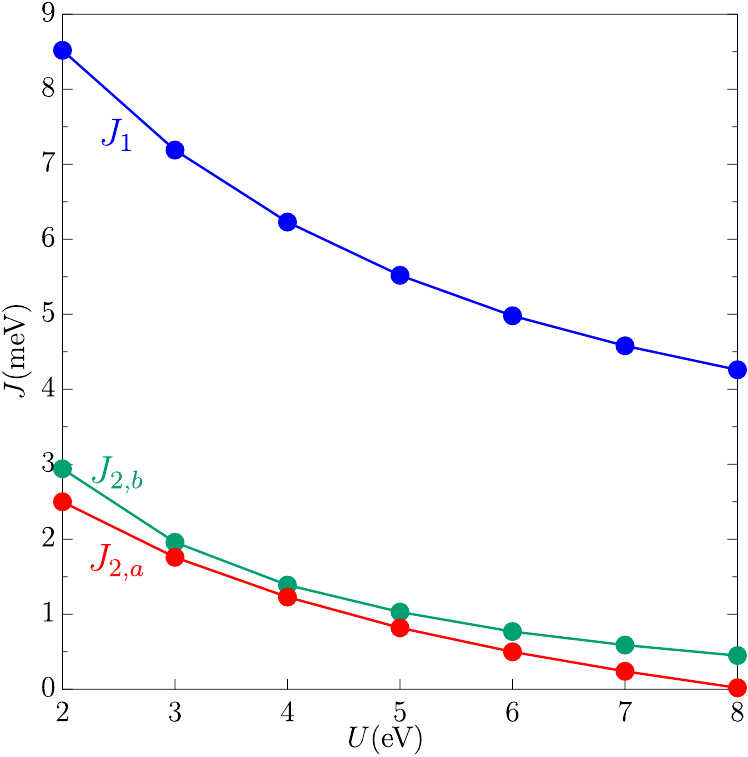}
    \caption{Estimated dominating isotropic Heisenberg exchanges $J_1$, $J_{2,a}$, $J_{2,b}$ via the TEMA method. The calculations are performed for fixed Hund's coupling $J_{H}$ = 1 eV for a range of $U$ values. Longer range in-plane $J_i$s, as well as AFM $J_{\perp}$ can be found in Appendix~\hyperref[appx:extended_spin_model]{S4}.}
    \label{fig:TEMA_J123}
\end{figure}

The Heisenberg exchange parameters of Eq.(\ref{eq:iso_spin_model}) are obtained from 
TEMA~\cite{totalenergy_analysis0,total_analysis,total_analysis1}, as described in Sec.~\hyperref[sec:comp_det]{Computational Details}. The results are shown in Fig.~\ref{fig:TEMA_J123} as a function of the Hubbard $U$ used in DFT+$U$ calculations. 
 We find the NN exchange $J_1$ to be the largest and AFM. The NNN exchanges, $J_{2,a}$ and $J_{2,b}$, are also AFM, but smaller than $J_1$, hence the NN's preferred aligment will be AFM. The difference $J_2 = J_{2,a} - J_{2,b}\neq0$, albeit small, is what encodes the AM behaviour into the spin model.
 Overall, the decay of the exchanges with increasing electronic correlations $U$ is consistent with the simple Hubbard model formula, $J\propto t^2/U$. Given that the activation gap is best reproduced when $U=2$, we choose the corresponding TEMA results $J_1\!=\!8.52\text{ meV}$, $J_{2,a}\!=\!2.50\text{ meV}$, $J_{2,b}\!=\!2.94\text{ meV}$, as reasonable parameters for the material.
 These estimates are in good agreement with previous  reported results~\cite{koo2012analysis}.

We also calculated some longer range contributions, as shown in Appendix~\hyperref[appx:extended_spin_model]{S4}. As expected, the magnitude of further neighbor couplings dies out fast as a function of distance. In these extended calculations we also kept the first $J_{\perp}$ coupling, between Mn$_1$-Mn$_3$ and Mn$_2$-Mn$_4$, which is found to be very small but AFM, consistent with the G-AFM order. 

We note that Ref.~\cite{xu2012first} found  $J_1$ and $J_{2,a}$ to be AFM, but $J_{2,b}$ to be FM, with the magnitude hierarchy $J_1 \approx J_{2,a} \approx |J_{2,b}|/2$ by using a first principles approach. %Using our DFT energies with their reported approach and equations we do confirm our results, and do not obtain a FM $J_{2,b}$ coupling.
Using the same computational setup as in  Ref.~\cite{xu2012first}, we don't reproduce their results.
Moreover, their reported value is in contradiction with the GKA rules for a half-filled high-spin configuration as we discuss below.
%Interestingly, our values seem to be in contradiction to the GKA rules, from which one would expect a FM $J_{2,b}$ coupling.
%In the next section we explain the AFM origin of the computed values and provide a microscopic explanation for them.
%In what follows, we analyse the microscopic origin of the magnetic exchange couplings.

\subsection{Microscopic origin of the isotropic magnetic exchanges}

%There are three competing dominant AFM interactions between \(\text{Mn}^{2+}\) ions: the NN direct exchange interaction $\text{Mn}_1$-$\text{Mn}_2$ ($J_1$), that belong to different sub-lattices, the NNN superexchange interaction through the \(180\degree\) $\text{Mn}_1$-O-$\text{Mn}_1$ ($J_{2,a}$) and the NNN superexchange interaction through $\text{Mn}_1$-$\text{Se}_{a(b)}$-$\text{Mn}_1$ ($J_{2,b}$), with an angle of almost \(90\degree\). Each Mn experiences four NN interactions, and four NNN interactions, two via each channel (Mn-O-Mn and Mn-Se-Mn). 

%In Fig.~\ref{fig:microscopics}(a) we show a schematic sketch of the metal $d$-orbitals relative to the ligand $p$-orbitals. Due to the $D_{2h}$ crystal environment of Mn, the crystal field splits the $d$-orbital manifold completely, however, without forming particularly distinct groupings, with the split levels being relatively evenly spaced between them. In this picture, the ligands are  fully occupied, and the Hund's rule ensures five localized electrons per Mn the $S=5/2$ high spin state, avoiding double occupancy in the Hubbard sense. The Hubbard $U$ further increases the energy separation between the occupied and unoccupied states.
 
To explain the sign of the isotropic exchange interactions obtained by our first principles approach, we revisit the possible hopping processes described in the previous section.
The various contributions to the magnetic exchange can be written in terms of 
 virtual direct hoppings from Mn to Mn ($J_\gamma^{\text{dir}}$), or virtual indirect hoppings mediated by the ligands X=O, Se   ($J_\gamma^{\text{ind,O}}$, $J_\gamma^{\text{ind,Se}}$) (in DFT the same physics is described as lowering of the one-electron energy by hopping) as follows
\begin{equation}
    J_\gamma = J_\gamma^{\text{dir}} + J_\gamma^{\text{ind,O}} + J_\gamma^{\text{ind,Se}},
\end{equation}
where $\gamma$ represents a particular bond.

First, we discuss the processes which do not require intermediate states. %Thus, the picture we consider consists only of the Mn d orbitals and the ligands p orbitals.
%The exchange paths that include intermediate states will be discussed afterward.
%\im{If I were a referee I would get angry at this point. Either you do calculations within the Hubbard model and analyze it in terms of virtual hoppings or you do calculations in DFT and analyze them in terms of one-electron energy. But you cannot mix the two. In any event, I am going to shorten it dramatically, you do not need to explain the textbook stuff.}
%We start by applying the GKA rules~\cite{GoodenoughPR1955,goodenough1963magnetism} to  the $J_1$ bond. 
%There is only one possible virtual process that takes place in the direct Mn-Mn path ($J_1^{\rm dir}$) between the high spin half-filled d orbitals, where the spins on the $\rm Mn_1$ and $\rm Mn_2$ sites are aligned antiparallel, e.g. $\rm Mn_1(\uparrow)$ and $\rm Mn_2(\downarrow)$.
%An electron with spin up from the $\rm Mn_1$ is transmitted to the $\rm Mn_2$ with antiparallel aligned spins, thus making one d orbital on the $\rm Mn_2$ doubly occupied.
%The analogue process where both Mn sites are parallelly polarized is not possible because of the Pauli exclusion principle.
%This leads to a AFM coupling in the $J_1^{\rm dir}$ channel.
Since Mn in \ce{La2O3Mn2Se2} has a full spin-up $d$ shell (Fig.~\ref{fig:microscopics} (a)),  all hopping processes, whether between the like or unlike orbitals, contribute equally to the AFM coupling. In a DFT context, each of these hoppings pushes the occupied states down on the AFM case, but only lead to broadening (and no centre of gravity shift, in the lowest order) of the occupied band in the FM case. The most important contribution to $J_1^{\rm dir}$, therefore, will come from direct $dd\sigma$ hopping, albeit $dd\pi$ and $dd\delta$ also contributing to it.

The indirect Mn-X-Mn paths ($J_\gamma^{\text{ind,X}}$) involve assisted hopping via the fully occupied $p^6$ ligand states. Integrating out these states has two effects: one is the crystal field splitting, for instance between $t_{2g}$ and $e_g$ orbitals.
Since all five $d$-orbitals are occupied, it is of secondary importance here. Second, effective hoppings between the two metal sites appear, {\it wherever it is allowed by symmetry}, of the order of $t_{pd\nu}t_{pd\nu'}/\Delta_X$, where $\{\nu,\nu'\}=\sigma,\pi,\delta$. This effective hopping acts the same way as direct hopping, i.e., produces AFM coupling. Note that the often discussed superexchange FM interactions are either proportional to the Hund's coupling $J_H$ on the metal and appear in only partially-occupied shells (e.g., $d^3$), or are proportional to the Hund's rule coupling on the ligand, and have an additional small factor of $J_H(X)/\Delta_X$ compared to the AFM one.
%First, an electron from a double occupied p-orbital is transmitted to $\rm Mn_1$.
%Due to the high-spin configuration on the d-orbitals, only an electron with an opposite spin to the $d^5$ is allowed to hop.
%After this, the electron that remains in the half-filled p orbital has the same spin alignment as the $\rm Mn_1$ site.
%Next, there are two possible hopping processes to complete the exchange: $i)$ the hole on the p-orbital is filled by an electron from the $\rm Mn_2$ d-orbitals (charge transfer) or $ii)$ the remaining p electron hops to one of the $\rm Mn_2$ d orbitals (two hole process).
%In both options, the Pauli principle restricts the transmission to only electrons with opposite spin, which leads to AFM coupling, and excludes FM mechanism for the exchange.

%Up to this point, the above-mentioned mechanism is geometry independent and, hence, general for half filled metal sites. Therefore, the superexchange which includes electron transfer dominantly via a single ligand orbital or a direct hopping are AFM.
%This mechanism occurs in the 180° geometry in the $\text{NNN}_a$ bond, where the dominant virtual process will take place within the O $p_x$ orbitals, resulting in AFM $J_{2,a}$.
%The $\sim$~90° geometry in the $\text{NNN}_b$ bond has several virtual hopping paths which are possible which also include more than one Se-p orbital.

The largest possible indirect hopping is of the order of $t_{pd\sigma}^2/\Delta_X$ and is maximized when the bond angle is $180\degree$, as is in the case of the $\text{NNN}_a$ bond and O $p_x$ intermediate state. For the NN $J_1$ bond the angle is 90$^\circ$ and this process is not allowed, however, there are several allowed processes of the order of $t_{pd\sigma}t_{pd\pi}/\Delta_O$ that together provide sizeable effective hopping and there is an additional contribution via Se. Therefore, both $J_1$ and $J_{2,a}$ must be AFM.

For the  $\text{NNN}_b$ bond there are two Mn-Se-Mn paths, and the angle is $\sim 90^\circ$. This hopping is mostly $t_{pd\sigma}t_{pd\pi}/\Delta_{Se}$, but the two additive paths and the fact that $\Delta_{Se}<\Delta_O$ make $J_{2,b}^{\rm ind}$ similar in magnitude to $J_{2,a}^{\rm ind}$.

\section{Magnetism}\label{sec:magnetism}

Summarizing our results above, we find the NN Heisenberg coupling $J_1$ to be large and AFM, and the NNN $J_{2,a}$ and $J_{2,b}$ to be  AFM as well. For sufficiently large NNN exchanges the AFM pattern could be destabilized. At the mean field level, the AM order is stabilized if $|J_1|>|J_{2a}|+|J_{2b}|$, a condition that is very well satisfied in our computational result.
%Considering how close to the isotropic limit $J_{2,a}=J_{2,b}=J_2$ the system is, and that $J_2/J_1 \approx 0.2 $, the material is far from close to the classical transition point $J_2/J_1 = 0.5 $ out of the AFM ground state~\cite{chandra1988possible,dagotto1989phase,richter2010spin,gelfand1989zero,jiang2012spin,hu2013direct}. Generally, within the isotropic Heisenberg model, the AFM order will be stable when the effective coupling  $\widetilde{J} \!=\! J_1 \!-\! J_{2,a} \!-\! J_{2,b}$ is positive $\widetilde{J}>0$, as it is in this case. 
Moreover, while exceedingly small, we found the first out of plane exchange $J_{\perp}$  connecting top to bottom layer, to be also antiferromagnetic and resulting in the G-AFM order. 

\subsection{Magnetic anisotropy}
Experimentally, below $T_N \approx163~\text{K}$, \ce{La2O3Mn2Se2} orders in the G-AFM pattern, with local moments aligned along the $z$-axis. A very small total magnetization was observed, attributed to potential moment canting~\cite{Cava,Evans,xie2013magnetic}. It is worth noting, however, that the corresponding magnetic space group is I4'/mm'm, and the point group 4'/mm'm, which is not compatible with ferromagnetism \cite{Bilbao}, thus no interactions consistent with the crystal symmetry, such as bilinear or higher-order Dzyaloshinskii-Moriya terms \cite{IgorMnTe}
can cause any net canting. More likely, the FM signal comes from interstitial Mn and/or Mn vacancies.
%Our DFT calculations find a well-formed localized moment on the Mn sites, of spin length $S\!=\!5/2$. 
%The bottom Mn$_1$-Mn$_2$ square lattice layer admits a AFM pattern forming from $S=5/2$ moments. The top square lattice layer Mn$_3$-Mn$_4$ is related by primitive lattice translations to the bottom, i.e., $\mathbf{S}_{\text{Mn}_3} = \mathbf{S}_{\text{Mn}_2}$ and $\mathbf{S}_{\text{Mn}_4} = \mathbf{S}_{\text{Mn}_1}$. Together, they form the G-AFM pattern shown in Fig~\ref{fig:compafig}. Within the primitive unit cell, which is also the magnetic unit cell, we have two sub-lattices that are not inversion related, but $C_{4\uvec{a}_3}$ related, which classifies this compound as an altermagnet. 

%\subsection{The G-AFM order}

Since the Mn$^{2+}$ ion has zero orbital moment, the expected magnetic anisotropy that would pin the moment on the $z$-axis is exceedingly small, of the order of microelectronvolts per Mn, and therefore beyond the accuracy of DFT calculations (and may even include a non-negligible dipole-dipole contribution \cite{Prague}). 
%Neutron experiments on \ce{La2O3Mn2Se2} reported a magnetic anisotropy suggesting an easy $\uvec{a}_3$-axis (parallel to $z$-axisWith this in mind,
We performed GGA+SOC+$U$ calculations considering collinear magnetic configurations and found, as expected, that the DFT anisotropy energy is of the order of 10 ${\mu}eV$. DFT predicts an in-plane easy-axis anisotropy contrary to the experimentally reported $z$-axis anisotropy~\cite{Cava,Evans}. As mentioned above, the required accuracy for such a result is beyond DFT capabilities.

\subsection{Spin wave spectra}

\begin{figure}
    \centering
    \includegraphics[width=\linewidth]{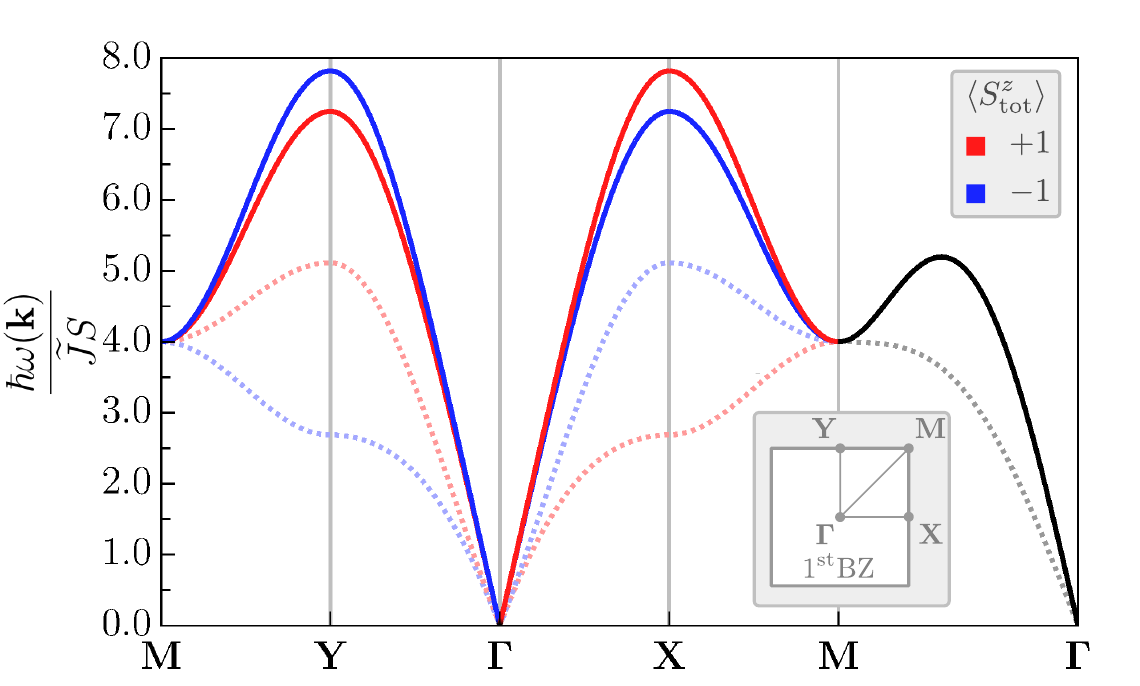}
    \caption{Spin $S^{z}_{\text{tot}}=+1,-1$ (red,blue) resolved magnon bands from LSWT, showing the characteristic AM splitting of chiral magnon bands. Black bands are doubly degenerate. The solid bands correspond to the $U\!=\!2\text{eV}$ TEMA result for the exchanges $J_1\!=\!8.52\text{meV}$, $J_{2,a}\!=\!2.50\text{meV}$, $J_{2,b}\!=\!2.94\text{meV}$, while, for comparison, the light dashed bands correspond to flipping the sign of $J_{2,b}\!=\!-2.50\text{meV}$. The magnon energy $\hbar\omega(\mathbf{k})$ is normalized to the effective AFM coupling $\widetilde{J} = J_1 - J_{2,a} - J_{2,b}$.}
    \label{fig:magnons}
\end{figure}

An important signature of an AM state is the presence of chiral magnons split along certain direction. However, the fact that our calculated $J_{2a}$ and $J_{2a}$ are, accidentally, rather close, suggests that this effect might be too small to be observed. This does not mean, however, that other signatures, such as exchange splitting of the one-electron bands, should be small. To make a quantitative assessment, we calculated the magnon bands within linear spin wave theory (LSWT), the details of which are shown in Appendix~\hyperref[appx:lswt]{S6}.

The chirality of the one magnon states can be identified by measuring the spin moment of the magnons. The Heisenberg model Eq.~(\ref{eq:iso_spin_model}) commutes with $S_{\text{tot}}^z$, the total spin moment along the quantization $\uvec{z}$-axis. As a result, non-degenerate one magnon states $|\psi_{\mu,\mathbf{k}}\rangle\!=\!\gamma^\dagger_{\mu,\mathbf{k}}|0\rangle$  will have a definite expectation value $\langle S^{z}_{\text{tot}}\rangle=\langle\psi_{\mu,\mathbf{k}}| S_{\text{tot}}^z |\psi_{\mu,\mathbf{k}}\rangle$. A finite expectation of $\langle S^{z}_{\text{tot}}\rangle$ imposes a handiness (chirality)  onto the magnons, with respect to the quantization along the $\uvec{z}$-axis.

In Fig.~\ref{fig:magnons} (solid lines) we plot the calculated magnon spectrum $\omega_{\mu,\mathbf{k}}$, for the TEMA $U\!=\!2~\text{eV}$ estimated exchanges, resolved by $\langle S^{z}_{\text{tot}}\rangle$. We also show, for comparison, the expected split magnon bands when considering  $J_{2,b}$ to be of the same magnitude but FM (dotted lines). While the conventional limit, $J_{2,a}=J_{2,b}$, would show all bands doubly degenerate, having $J_{2,a}\neq J_{2,b}$ results in split non-degenerate bands along specific paths, with the split bands acquiring a definite $\langle S^z_{\text{tot}} \rangle\!=\!\pm 1$. Additionally, the chiral magnon splitting is inverted around the $\mathbf{\Gamma}$ point, for example along $\mathbf{Y}\shyph\mathbf{\Gamma}$ vs $\mathbf{\Gamma}\shyph\mathbf{X}$, owing to the $C_{4\uvec{a}_3}$ magnetic sub-lattice relation. This is a characteristic signature of AM state, which at the spin model level is a result of the difference between $J_{2,a}$ and $J_{2,b}$. 
As expected, the chiral splitting $\Delta\omega(\mathbf{X})$ is fairly small and is maximized at the $\mathbf{X}$- and $\mathbf{Y}$- point. Comparatively, if $J_{2,b}$ where FM the splitting is much more pronounced (Fig.~\ref{fig:magnons} (dotted lines)). In either case, the magnon splitting is found to be
%To illustrate that this is a consequence of the accidental near-degeneracy of  $J_{2,b}$ and $J_{2,b}$,
%we make a ``gedanken experiment'' wherein we flip the sign of the $J_{2,b}$ without changing its magnitude. In that case,  we observe a huge magnon band splitting, as shown in  Fig.~\ref{fig:magnons} as well. 
%The characteristic shape of the magnon bands can be used as a measure for determining the sign of the $J_{2,b}$ exchange.  
\begin{equation}
    \hbar\Delta\omega(\mathbf{X}) = 4S\sqrt{\left|J_{2,a}-J_{2,b}\right|\left(2J_1-J_{2,a}-J_{2,b}\right)},
\end{equation}
which sets a direct connection between the observation of  $\Delta\omega(\mathbf{X})$ from inelastic neutron scattering experiments and the exchange parameters $J_1$, $J_{2,a}$ and $J_{2,b}$.

There is an additional feature along the $\mathbf{M}\shyph\mathbf{\Gamma}$ line (where the chiral splitting is absent by symmetry), namely a local maximum, which appears to be parameter-dependent. Given $J_1\!>\!0$ and $J_1 -  J_{2,a} -J_{2,b} \!>\! 0$ to guarantee the collinear AFM state as the ground state of the spin model, we find that the maximum  appears when $J_{2,b}\!>\!0$ and $-J_{2,b}\!<\!J_{2,a}$ at the $\mathbf{k}_o$ momentum point
\begin{equation}
\begin{array}{c}
    \mathbf{k}_o \!=\! \dfrac{f_o}{2}(\mathbf{b}_1\!+\!\mathbf{b}_2),\\[12pt]
    f_o\!=\!\dfrac{1}{\pi} \cos^{-1}\!\left(-\dfrac{J_1 \!-\!  J_{2,a} \!-\! J_{2,b}}{J_1 \!+\!  J_{2,a} \!+\! J_{2,b}}\right),
\end{array}
\end{equation}
where $\mathbf{b}_i$ are the reciprocal lattice vectors. In the extreme case of $J_{2,a},J_{2,b}\xrightarrow{}J_1\!>\!0$, on the brink of stability of the AFM phase, the local maximum lies exactly in the middle of the $\mathbf{M}\shyph\mathbf{\Gamma}$ path $(f_o=1/2)$, with the $\mathbf{M}$ point being gapless, signalling the end of the stability of the AFM phase. From this extremum parameter point, consider reducing the value of $J_{2,a}$, such that $0\!<\!J_{2,a}\!<\!J_{2,b}$. Then the $\mathbf{M}$ point becomes gapped and grows in value, while the maximum starts to soften and moves towards the $\mathbf{M}$ point $(1\!>\!f_o\!>\!1/2)$. Considering negative exchange parameters,  the more we negatively grow $J_{2,a}<0$, the closer to the $\mathbf{M}$ point the maximum gets $(1\!\gtrsim\!f_o)$. Finally, if we also consider a negative $J_{2,b}$ with $J_{2,a},J_{2,b}\!<\!0$,  the maximum collapses at the $\mathbf{M}$ point $(f_o=1)$, and becomes an immovable saddle point on the Brillouin zone edge. This feature may thus be used as another indicator of whether $J_{2,a}$ and $J_{2,b}$ are of the same or of different signs.

%where $\mathbf{b}_i$ are the reciprocal lattice vectors. In the extreme case of $J_{2,a},J_{2,b}\xrightarrow{}J_1\!>\!0$, on the brink of stability of the AFM phase, the local maximum lies exactly in the middle $f_o=1/2$ of the $\mathbf{M}\shyph\mathbf{\Gamma}$ path, and the 
%\rv{ could you revise the following sentences, it is twice mentioned that the M point becomes gapless, the flow is confusing} $\mathbf{M}$ point  becomes gapless, signalling the end of the stability of the AFM phase. Slowly moving backwards from there by reducing $0<J_{2,a}\!<\!J_{2,b}$, the $\mathbf{M}$ point becomes gapless and grows in magnitude, making the maximum slowly soften up, and move in the direction $f_o\!>\!1/2$, i.e., towards the $\mathbf{M}$ point. By the time we switch into the negatives $J_{2,a}\!<\!0$, the maximum $1>f_o\gg1/2$ is approaching the $\mathbf{M}$ point significantly, and is becoming barely visible. Once $J_{2,a}\!<\!-J_{2,b}$ the maximum $f_o=1$ coincides with the $\mathbf{M}$ point. If we drop the conditions $J_{2,b}\!>\!0$ and $-J_{2,b}\!<\!J_{2,a}$, and enter the case of $J_{2,a}\!<\!0,\ J_{2,b}\!<\!0$ (both are FM), the maximum coincides at the $\mathbf{M}$ point. This feature may thus be used as a good indicator of whether $J_{2,a}$ and $J_{2,b}$ are of the same or of different signs.

\section{Conclusions}
%\pps{Maybe expand and beautify the conclusions?}
%\rv{we can finalize this in the second round}

In summary, we have investigated the origin of the exchange interactions and their signatures in the magnetic excitations of the  high-spin $S=5/2$ Mn-based AM insulator \ce{La2O3Mn2Se2}, which shows a G-AFM order. We confirmed with previous literature that DFT gives a good description of its electronic and magnetic properties.
%Owing to the $C_4$ relation, but not inversion, of the magnetic sub-lattices, this compound is classified as an insulating AM within the Lieb lattice. 
By performing a Wannierization analysis of the electronic structure we have been able to decipher the most dominant interactions. We find that both the direct Mn-Mn $d_{xy}-d_{xy}$ interaction and the Mn-Mn $d_{x^2-y^2}-d_{xy}$ superexchange dominate the nearest neighbor $J_1$ magnetic exchange.
%Extracting Heisenberg exchanges using TEMA, we found a large AFM $J_1$ exchange, which originates from the two dominant hoppings just mentioned.
%both from the direct NN $d_{xy}$ orbital hopping, but also the Mn-O-Mn superexchange path coming from the \(d_{x^2-y^2}\) orbitals. 
Contrary to earlier conjectures, it is the latter contribution that determines the strength of $J_1$. 
%Both the direct and the indirect Mn-O-Mn hopping contribute AFM to the total $J_1$, the former explained from the standard kinetic argument and the latter as expected from the GKA rules. 
The NNN couplings $J_{2,a}$ and $J_{2,b}$ were found to be AFM, and, consistent with the GKA rules of a high spin configuration system with all five orbitals occupied in the spin-majority channel.
%going against some past literature but consistent with the GKA rules of a high spin configuration system.

The accidental similar values of  $J_{2,a}$ and $J_{2,b}$ suppress, but not fully eliminate the chiral splitting of the magnons. Other materials of the same family are likely to lack this similarity, which would make them ideal platforms for observing chiral magnons. We have identified general signatures in the magnon bands, which can help assess the isotropic and anisotropic exchanges directly from experiments. The signs of $J_{2,a}$ and $J_{2,b}$ are found to correlate with the overall magnon shape around $\mathbf{X}$, $\mathbf{Y}$ points, as well as the presence or absence of a local maximum along the $\mathbf{\Gamma}\shyph\mathbf{M}$ line. The $\mathbf{X}/\mathbf{Y}$ point splitting is related to the isotropic maximum chiral splitting in the AM state, with inverts the chiral magnons between the two points owing to the $C_4$ sub-lattice relation, while the $\mathbf{M}$ point is doubly degenerate. However, anisotropic effects will split the magnons at the $\mathbf{M}$ point. Although expected to be small in $3d$-systems, the magnitude of anisotropic exchanges can be observed and constrained from the $\mathbf{M}$ point splitting experimentally, even if the spin gap is difficult to observe.

Finaly, previously discussed potential spin canting \cite{Cava,xie2013magnetic} is not allowed by symmetry in the magnetic point group 4'/mm'm, corresponding to the easy-axis anisotropy, albeit it would have been  allowed in the m'm'm point group, corresponding to the easy-plane anisotropy. In the latter case one would expect some canting, for instance, because of  the in-plane easy direction rotating between the two sublattices.

%within $C_4$ symmetric AM candidates, we were able to make some general statements about the possibility, and origin, of canting. If the system is easy-axis, there will be no canting from spin bilinear exchanges, as checked from the symmetry reduced anisotropic model. However, easy-plane magnets will naturally cant, but not due to conventional DM exchange (which completely cancels out), but rather due to a permuted SIA between the two sublattices. This should be a general feature present in any $C_4$ symmetric AM candidate with magnetic moments larger than 1/2.

%\section{Bullet points for conclusions (to be removed, use only as a guide)}
%\begin{itemize}
%    \item via ab initio analysis methods we find pure AFM $J_{1}$,$J_{2,a}$, and $J_{2,b}$
%    \item this is in contracting by previous reports and expectations
%    \item the surprising AFM exchange can be microscopically explained by ...
%    \item FM $J_{2,b}$ vs AFM $J_{2,b}$ should be possible to be directly seen in low temperature neutron scattering by a comparison of the magnon band splitting size.
%    \item observed tiny spin canting not in the bilinears spin model. DM interactions cancel out. SIA may cause canting in $C_4$ symmetric AM generically but not in our case. These should be fairly universal signatures.
%    \item Signatures of isotropic and anisotropic parameters in the magnon spectrum. Again these should be fairly universal signatures. Inelastic neutrons proposed to set them experimentally from our analysis.
%\end{itemize}

\section{Acknowledgements}
L.G-G., A.R., P.P.S and R.V. thank the Deutsche Forschungsgemeinschaft (DFG, German Research Foundation) through the TRR 288 - 422213477 (project B05) and project Nr. VA 117/23-1 — 509751747. 
 I.I.M. was supported by the Army Research Office under Cooperative Agreement Number W911NF-22-2-0173. He also acknowledges the WE Heraeus Foundation for supporting his visits at the Goethe University Frankfurt. The structural unit cell of \ce{La2O3Mn2Se2} was presented using the software VESTA \cite{VESTA}.

\section{Data availability}
The datasets generated during the current study are available upon reasonable request.

\bibliography{bibliography}

\subfile{SM/Supplementary_material}

\end{document}

%% file: SM/Supplementary_material.tex
%\newpage
%\clearpage
%\beginsupplement
\appendix

\section{S1: Symmetry}\label{appx:symm_anal}

%\section{Crystal Symmetry}\label{subsec:crystal_symm}

%The crystal structure of the layered compound   in the conventional unit cell $\mathbf{a}_1$, $\mathbf{a}_2$, $\mathbf{a}_3$ is shown in the main text Fig.~\ref{fig:compafig}(a). The [Mn$_2$OSe$_2$]$^{2-}$ layers, with \ce{Mn} shown in blue, \ce{O} in red, and \ce{Se} in green, are separated by sheets of [La$_2$O$_2$]$^{2+}$, with \ce{La} shown in grey.

\ce{La2O3Mn2Se2} respects the \(I4/mmm\) space group symmetry (number 139), with the associated crystal point group \(D_{4h}\). It consists of 16 symmorphic point group operations,   
\begin{equation}
    \begin{array}{llll}
        E,                    &  C_{2\uvec{a}_1},                  & I,                    & m_{\uvec{a}_1},              \\
        C_{4\uvec{a}_3},      &  C_{2\uvec{a}_2},                  & S_{4\uvec{a}_3},      & m_{\uvec{a}_2},              \\
        C^{-1}_{4\uvec{a}_3}, &  C_{2(\uvec{a}_1\!+\uvec{a}_2)},   & S_{4\uvec{a}_3}^{-1}, & m_{\uvec{a}_1\!+\uvec{a}_2}, \\
        C_{2\uvec{a}_3},      &  C_{2(\uvec{a}_1\!-\uvec{a}_2)},   & m_{\uvec{a}_3},       & m_{\uvec{a}_1\!-\uvec{a}_2},
    \end{array}
\end{equation}
where $E$ the identity, $I$ the inversion, $C_{\alpha\uvec{n}}$ the $\alpha$-fold rotation around $\uvec{n}$-axis, $m_{\uvec{n}}=IC_{\alpha\uvec{n}}$ a mirror with $\uvec{n}$ perpendicular to the mirror plane, and $S_{4\uvec{n}} = m_{\uvec{n}}C_{4\uvec{n}}$ the $\alpha$-fold improper rotation. 

Along with the point group elements the space group also has all the primitive lattice translations, with the primitive lattice vectors $\mathbf{a}_i^p$ that define them following from the conventional lattice vectors $\mathbf{a}_i$ in the standard way
\begin{equation}
    \left(\begin{array}{l}
         \mathbf{a}_1^p \\[4pt]
         \mathbf{a}_2^p \\[4pt]
         \mathbf{a}_3^p  
    \end{array}\right)
=
    \left(\begin{array}{rrr}
         -\frac{1}{2} & \frac{1}{2} & \frac{1}{2} \\[4pt]
          \frac{1}{2} & -\frac{1}{2} & \frac{1}{2} \\[4pt]
          \frac{1}{2} & \frac{1}{2} & -\frac{1}{2} 
    \end{array}\right)    
    \left(\begin{array}{l}
         \mathbf{a}_1 \\[4pt]
         \mathbf{a}_2 \\[4pt]
         \mathbf{a}_3  
    \end{array}\right).
\end{equation}

Model terms can be symmetry reduced under the self-mapping subgroup, i.e. the subset of space group operations that maps one site onto its self, or one bond onto its self. We now analyze these subgroups.

\subsection{On-site symmetry}

The metal Mn sites occupy the Wyckoff position 4c, which admits a $D_{2h}$ point group, a proper subgroup of $D_{4h}$. Considering the first sub-lattice \ce{Mn1} the self-mapping subgroup explicitly reads
\begin{equation}\label{eq:onsite_sym}
    D_{2h}  \cong \{ E,\ C_{2\uvec{a}_3},\ C_{2\uvec{a}_1},\ C_{2\uvec{a}_2},\ I,\ m_{\uvec{a}_3},\ m_{\uvec{a}_1},\ m_{\uvec{a}_2} \},
\end{equation}
up to lattice translations. The excluded operations of the space group will generate the second sub-lattice \ce{Mn2}, which is thus symmetry related to the first and not independent. Note that the inversion symmetry $I$ is a self-mapping operation of \ce{Mn1}, and does not relate one sublattice to the other. Sub-lattices Mn$_3$ and Mn$_4$ in the conventional cell are related by primitive lattice translations to Mn$_2$ and Mn$_1$ respectively. 

\subsection{Bond symmetry}\label{sec:bond_sym}

While for the on-site symmetry we are afforded the ease of looking up the Wykoff position from crystallographic tables and immediately retrieving the self-mapping subgroup Eq.~(\ref{eq:onsite_sym}), when considering the bonds a bit more work is required. The task is to check which of the operations of $D_{4h}$ will leave the bonds invariant, up to lattice translations. Additionally, having two sites, $i$ and $j$, per bond, we need to also include operations that map bond $(i\shyph j)$ to bond $(j\shyph i)$, since they are related simply by a transposition. Bond subgroup symmetry element $g$ that require this additional transposition will be indicated as $g^T$. The bonds under consideration are shown in the main text Fig.~\ref{fig:compafig}.

\subsubsection{NN bond}
We start with the NN bond where upon applying all the space group symmetries we find that the self-mapping subgroup consists of
\begin{equation} \label{eq:B1_sym}
C_{2v}  \cong \{ E,\ C_{2(\uvec{a}_1\!-\uvec{a}_2)}^T,\ m_{\hat{\mathbf{a}}_3}^T,\ m_{\hat{\mathbf{a}}_1\!+\hat{\mathbf{a}}_2} \},
\end{equation}
up to lattice translations. The collection of operations is isomorphic to a $C_{2v}$ group, which is a subgroup of $D_{4h}$. 

\subsubsection{NNN bond}

For the NNN bonds we find that, even through they have exactly the same length, bond NNN\textsubscript{$a$} can never be mapped onto bond  NNN\textsubscript{$b$} by any of the space group operations. Specifically we find that for bond NNN\textsubscript{$b$} the self-mapping operations are
\begin{equation}\label{eq:B2_sym}
\begin{array}{rl}
    D_{2h}  \cong  &\{           E,\   C_{2\hat{\mathbf{a}}_3}^T,\  C_{2\hat{\mathbf{a}}_1}^T,\ C_{2\hat{\mathbf{a}}_2}, \\[6pt]
                   &\phantom{\{} \ \ I^T,\ m_{\hat{\mathbf{a}}_3},\ m_{\hat{\mathbf{a}}_1}\     m_{\hat{\mathbf{a}}_2}^T, \},
\end{array}
\end{equation}
up to lattice translations, and for bond NNN\textsubscript{$a$} we similarly have,
\begin{equation}\label{eq:B3_sym}
\begin{array}{rl}
    D_{2h}  \cong  &\{           E,\   C_{2\hat{\mathbf{a}}_3}^T,\  C_{2\hat{\mathbf{a}}_1},\   C_{2\hat{\mathbf{a}}_2}^T, \\[6pt]
                   &\phantom{\{} \ \ I^T,\ m_{\hat{\mathbf{a}}_3},\ m_{\hat{\mathbf{a}}_1}^T,\  m_{\hat{\mathbf{a}}_2} \},
\end{array}
\end{equation}
up to lattice translations. Both NNN\textsubscript{$a$} and NNN\textsubscript{$b$} are invariant under the $D_{2h}$ group. This is not supersizing in the sense that the sub-lattice self-mapping subgroup is $D_{2h}$, and for NNN bonds both bond sites belong to the same sub-lattice.

\section{S2: Slater-Koster approximation} \phantomsection\label{appx:sk}

In the SK analysis \cite{SKanalysis1954} we make the approximation of using only 
two central integrals in the estimation of the tight-binding parameter 
$t_{ij} = \int \mathrm{d}\mathbf{r} \psi_i(\mathbf{r}) H(\mathbf{r}) \psi_j(\mathbf{r}) \simeq \int\mathrm{d}\mathbf{r} \psi_i(\mathbf{r}) H_{ij}(\mathbf{r}) \psi_j(\mathbf{r}),$
where $\psi_{i(j)}(\mathbf{r})$ the orbital on site $i$($j$), 
$H(\mathbf{r})$ the full Hamiltonian with the entire crystal potential generated by every site, and $H_{ij}(\mathbf{r})$ the Hamiltonian involving only site $i$ and $j$ potentials. From a symmetry perspective, this is equivalent to treating every bond as an isolated diatomic molecule. Therefore, it is equivalent to imposing $C_{\infty v}$ symmetry for heterogeneous bonds, and $D_{\infty v}$ symmetry for homogeneous bonds. The crystal symmetry is recovered when considering the virtual processes of metal to ligand, which effectively incorporate the true symmetry geometrically in the virtual paths. In the following, we write down the primary SK parameters assuming $z$-axis hoppings, define the rotations that put us on the desired crystal bond, and derive the final SK reduced hopping matrices.

\subsection{Primary SK hopping parameters}
Consider the heterogeneous metal-to-ligand bonds, with the Mn $d$-orbitals at the origin, and the ligand $p$-orbitals along the $\uvec{z}$-axis.  The $d\shyph p$ matrix, which respects  $C_{\infty v}$ symmetry, is given as
\begin{equation}
    T_{d\shyph p,X}^{(\hat{\mathbf{z}})} \!=\! 
    \begin{matrix}
        \begin{array}{rrr}
            \makebox[\widthof{$t_{X,dp\pi}$}][r]{${p_x}$} &
            \makebox[\widthof{$t_{X,dp\pi}$}][r]{${p_y}$} & 
            \makebox[\widthof{$t_{X,dp\sigma}$}][r]{${p_z}$} 
        \end{array} &
        \begin{matrix} 
            \makebox[\widthof{${x^2\!\!-\!y^2}$}]{\phantom{0}} 
        \end{matrix} \\[4pt]
        \left(
        \begin{array}{rrr}
                      0 &           0 &              0 \\
                      0 &           0 & t_{X,dp\sigma} \\
                      0 & t_{X,dp\pi} &              0 \\       
            t_{X,dp\pi} &           0 &              0 \\
                      0 &           0 &              0 \\
        \end{array}
        \right) &
        \begingroup
            \arraycolsep=-18pt
            \begin{array}{l} 
                {x^2\!\!-\!y^2} \\ {z^2} \\ {yz} \\ {zx} \\ {xy} 
            \end{array}
        \endgroup
    \end{matrix}
\end{equation}
for ligand $X$ = Se, O. The homogeneous bond between Mn $d$-orbitals, placed on the $\hat{\mathbf{z}}$ axis, is
%\begin{equation}\label{eq:sk_z_dd_hop}
%    T_{d\shyph d}^{(\hat{\mathbf{z}})} = 
%    \left(\begin{array}{ccccc}
%        t_{\delta} & 0 &  0 &  0 & 0 \\
%        0 &  t_{\sigma} & 0 & 0 & 0 \\
%        0 &  0 & t_{\pi} & 0 & 0  \\
%        0 &  0 & 0 & t_{\pi} & 0 \\
%        0 &  0 & 0 & 0 &  t_{\delta} \\
%    \end{array}
%    \right),
%\end{equation}
%
%
\begin{equation}\label{eq:sk_z_dd_hop}
    \begin{array}{l}
        T^{(\hat{\mathbf{z}})}_{d\shyph d} = 
        \begin{matrix}
            \begingroup
                \arraycolsep=2pt
                \begin{array}{rrrrr}
                    \makebox[\widthof{${x^2\!\!-\!y^2}$}][r]{${x^2\!\!-\!y^2}$}  &
                    \makebox[\widthof{$t_{dd\sigma}$}][r]{${z^2}$} & 
                    \makebox[\widthof{$t_{dd\pi}$}][r]{${yz}$} & 
                    \makebox[\widthof{$t_{dd\pi}$}][r]{${zx}$} & 
                    \makebox[\widthof{$t_{dd\delta}$}][r]{${xy}$}
                \end{array} 
             \endgroup &
            \begin{matrix} 
                \makebox[\widthof{${x^2\!\!-\!y^2}$}]{\phantom{0}} 
            \end{matrix} \\[4pt]
            \left(
            \begingroup
                \arraycolsep=2pt
                \begin{array}{rrrrr}
                    \makebox[\widthof{${x^2\!\!-\!y^2}$}][r]{$t_{dd\delta}$}
                               &            0 &         0 &         0 &             0 \\
                             0 & t_{dd\sigma} &         0 &         0 &             0 \\
                             0 &            0 & t_{dd\pi} &         0 &             0 \\
                             0 &            0 &         0 & t_{dd\pi} &             0 \\
                             0 &            0 &         0 &         0 &  t_{dd\delta} \\
                \end{array}
            \endgroup
            \right) &
            \begingroup
                \arraycolsep=-18pt
                \begin{array}{l} 
                    {x^2\!\!-\!y^2} \\ {z^2} \\ {yz} \\ {zx} \\ {xy} 
                \end{array}
            \endgroup
        \end{matrix}
    \end{array}
\end{equation}
 which respects the $D_{\infty v}$ symmetry. The sign of the sigma integrals $t_{dd\sigma}$, $t_{X,dp\sigma}$, pi integrals $t_{dd\pi}$, $t_{X,dp\pi}$, and the delta integral $t_{dd\delta}$ cannot be determined by symmetry. The physical expectation about which sign orbital lobes overlap in the hopping integral lead to the expected signs $t_{\sigma},t_{\delta}, t_{X,\pi}<0$ and $t_{\pi}, t_{X,\sigma}>0$.

\subsection{SK direct hopping matrix by rotation}

\begin{table}[t!]
    \centering
    \begin{tabular}{c r c | c r c}
    \hline
    \hline
    bond &  $\mathbf{n}$  & $\ \ \ \ \phi \ \ \ \ $ & bond & $\mathbf{n}$ & $\ \ \ \ \ \ \phi \ \ \ \ \ \ $ \\[2pt]
    \hline
    \(\mathrm{Mn}_1\shyph\mathrm{Mn}_2 \)    &  \(-\hat{\mathbf{x}}+\hat{\mathbf{y}}\) &  \(\pi/2\) &
    \(\mathrm{Mn}_1\shyph\mathrm{Se}_a \)    &  \(-\hat{\mathbf{x}}\) &  \(\pi/4 + \theta\) \\[2pt] 
    \(\mathrm{Mn}_1\shyph\mathrm{Mn}_1' \)   &  \(\hat{\mathbf{y}}\) &  \(\pi/2\) &
    \(\mathrm{Mn}_2\shyph\mathrm{Se}_a\)     &  \(-\hat{\mathbf{y}}\) &  \(\pi/4 + \theta\) \\[2pt]
    \(\mathrm{Mn}_1\shyph\mathrm{Mn}_1''\)   &  \(-\hat{\mathbf{x}}\) &  \(\pi/2\) &
    \(\mathrm{Mn}''_1\shyph\mathrm{Se}_a\)   &   \(\hat{\mathbf{x}}\) &  \(\pi/4 + \theta\) \\[2pt]
    \(\mathrm{Mn}_1\shyph\mathrm{O}\)        &  \(\hat{\mathbf{y}}\) & \(\pi/2\) &
    \(\mathrm{Mn}_1\shyph\mathrm{Se}_b\)     &  \(-\hat{\mathbf{x}}\) &  \(3\pi/4 - \theta\) \\[2pt]
    \(\mathrm{Mn}_2\shyph\mathrm{O}\)        &  \(\hat{\mathbf{x}}\) &  \(\pi/2\) &
    \(\mathrm{Mn}_2\shyph\mathrm{Se}_b\)     &  \(-\hat{\mathbf{y}}\) & \( 3\pi/4 - \theta\) \\[2pt]
    \(\mathrm{Mn}'_1\shyph\mathrm{O}\)       &  \(-\hat{\mathbf{y}}\) &  \(\pi/2\) &
    \(\mathrm{Mn}''_1\shyph\mathrm{Se}_b\)   &  \( \hat{\mathbf{x}}\) &  \(3\pi/4 - \theta\) \\
    \hline
    \hline
    \end{tabular}
    \caption{Axis $\mathbf{n}$ and angle $\phi$ of rotation, in the global basis, to rotate SK bonds from the $\uvec{z}$-axis onto the crystal bond. The angle $\theta = 47.7\degree-45\degree = 2.7\degree$ in the crystal structural. Due to its negligible difference from the ideal angle, we report SK matrices for $\theta = 45\degree$.}\label{tab:bond_angle_axis_for_sl}
\end{table}

To bring the primary SK matrix $T^{(\uvec{z})}$ into the crystal bond $(i\shyph j)$ SK matrix $T_{i\shyph j}$, we perform an active rotation around axis $\mathbf{n}_{ij}$ with angle $\phi_{ij}$, as listed in Table~\ref{tab:bond_angle_axis_for_sl} for all bonds needed. The orbital rotation is then performed as
\begin{equation}\label{eq:rot_op}
     T_{i\shyph j} = U_{o(i)}(\phi_{ij},\mathbf{n}_{ij})\ T_{o(i)\shyph o(j)}^{(\uvec{z})} U_{o(j)}^{\dagger}(\phi_{ij},\mathbf{n}_{ij})
\end{equation}
where $o(i),\ o(j)\!=\!p\text{ or }d$ the orbitals belonging to bond sites $i$, $j$. The orbital rotation matrix are constructed as
\begin{equation}
    U_{p(d)}(\phi,\mathbf{n}) = \mathrm{e}^{-i\phi\mathbf{n}\cdot\mathbf{L}^{p(d)}}
\end{equation}
where $\mathbf{L}^{p(d)}$ are the angular momentum $l=1$($2$) matrices, written in the orbital basis $d$($p$), following the same orbital order as in the definition of $T^{(\uvec{z})}$. Finally, when $j\!=\!\mathrm{Mn}_2$, we need to bring the sublattice into its local basis 
\begin{equation}
    T_{i\shyph \text{Mn}_2} \rightarrow T_{i\shyph \text{Mn}_2} U_{\text{local}}, 
\end{equation}
by performing the passive local basis transformation
\begin{equation}
    U_{\text{local}}=
    \left(
    \begin{array}{rrrrr}
       -1  &  0  &  0  &  0  &  0  \\
        0  &  \phantom{-}1  &  0  &  0  &  0  \\
        0  &  0  &  0  & -1  &  0  \\
        0  &  0  &  \phantom{-}1  &  0  &  0  \\
        0  &  0  &  0  &  0  & -1  \\
    \end{array}
    \right).
\end{equation}
This completes the construction of the SK matrices $T_{i\shyph j}$ in the local basis used in the main text.

\subsubsection{Direct metal to metal SK matrix}

For the NN bonds, using the aforementioned procedure, we find the direct SK matrix for NN
\begin{equation}
    T_{\text{Mn}_1\shyph \text{Mn}_2} =
    \left(
    \begin{array}{cc}
        \mathbb{M}_{e}    & \mathbb{M}_{et} \\ 
        \mathbb{M}_{et}^T & \mathbb{M}_{t}
    \end{array}
    \right),     
\end{equation}
and for NNN
\begin{equation}
    T_{\text{Mn}_1\shyph \text{Mn}_1'} =
    \left(
    \begin{array}{cc}
        \mathbb{M}'_{e}           & \mathbb{0}_{2\!\times\!3} \\ 
        \mathbb{0}_{3\!\times\!2} & \mathbb{M}'_{t}
    \end{array}
    \right),     
\end{equation}
reported in block of $e_g-e_g$ (e), $e_g-t_{2g}$ (et) and $t_{2g}-t_{2g}$ (t) sectors. The blocks are as follows
\begin{equation}
    \mathbb{M}_{e} =
    \left(
    \begin{array}{cc}
        -t_{dd\pi} &                                    0 \\[6pt]
                0  & \dfrac{t_{dd\sigma}+3t_{dd\delta}}{4}
    \end{array}
    \right)     
\end{equation}
\begin{equation}
\mathbb{M}_{et} =
\left(
\begin{array}{ccc}
 \ \ \ 0 \ \ \ & \ \ \ 0 \ \ \  & 0 \\[6pt]
 \ \ \ 0 \ \ \  & \ \ \ 0 \ \ \  & \dfrac{\sqrt{3}(t_{dd\sigma}-t_{dd\delta})}{4} 
\end{array}
\right),     
\end{equation}
\begin{equation}
    \begin{array}{l}
    \mathbb{M}_{t} = \\[4pt]
    \left(
    \begin{array}{ccc}
        \dfrac{t_{dd\pi} - t_{dd\delta}}{2} & -\dfrac{t_{dd\pi} + t_{dd\delta}}{2} & 0 \\[4pt]
        \dfrac{t_{dd\pi} + t_{dd\delta}}{2} & \dfrac{-t_{dd\pi} + t_{dd\delta}}{2} & 0 \\[6pt]
                                          0 &                                    0 & -\dfrac{3t_{dd\sigma}+t_{dd\delta}}{4}
    \end{array}
    \right),     
    \end{array}
\end{equation}
\begin{equation}
    \mathbb{M}'_{e} =
    \left(
    \begin{array}{cc}
                \dfrac{3t'_{dd\sigma}+t'_{dd\delta}}{4} & \dfrac{\sqrt{3}t'_{dd\sigma}-t'_{dd\delta}}{4} \\[6pt]
        \dfrac{\sqrt{3}t'_{dd\sigma}-t'_{dd\delta}}{4} & \dfrac{t'_{dd\sigma}+3t'_{dd\delta}}{4}
    \end{array}
    \right),     
\end{equation}
\begin{equation}
    \mathbb{M}'_{t} =
    \left(
    \begin{array}{ccc}
        t'_{dd\pi} &             0 &             0 \\
                 0 & t'_{dd\delta} &             0 \\
                 0 &             0 & t'_{dd\sigma} 
    \end{array}
    \right).     
\end{equation}
Note how the matrix has a higher symmetry than the expected symmetry discussed in Appendix~\hyperref[appx:additional_dft_res]{S3}, and that $\text{NNN}_{a}$ and $\text{NNN}_{b}$ both have the same SK parameters $t'$. This is all due to the artificially enhanced symmetry imposed by the SK approximation as discussed at the beginning of this section. When the ligands are integrated out perturbatively, the effective hoppings recovered will restore the appropriate symmetry.

\subsubsection{Metal to ligand hopping}

For the metal-to-ligand bonds, using the aforementioned procedure, we find the direct SK matrices
\begin{equation}
    T_{\mathrm{Mn}_1\shyph\mathrm{O}}   =  
    \left(
    \begin{array}{ccc}
     \dfrac{\sqrt{3}t_{\mathrm{O},dp\sigma}}{2}   & 0 & 0 \\
     -\dfrac{t_{\mathrm{O},dp\sigma}}{2} & 0 & 0 \\
     0 & 0 & 0 \\
     0 & 0 & t_{\mathrm{O},dp\pi} \\
     0 & t_{\mathrm{O},dp\pi} & 0 \\
    \end{array}
    \right),
\end{equation}
and using the ideal Se angle of $45\degree$ (since the difference with the $47.7\degree$ is negligible)
\begin{equation}
    \begin{array}{l}
        T_{\mathrm{Mn}_1\shyph\mathrm{Se_a}}   =     \\[4pt]
        \left(
        \begingroup
            \arraycolsep=2pt
            \begin{array}{ccc}
                 0 & 
                 \frac{-\sqrt{3} t_{\text{Se},dp\sigma}-2 t_{\text{Se},dp\pi}}{4\sqrt{2}} & 
                 \frac{-\sqrt{3} t_{\text{Se},dp\sigma}+2 t_{\text{Se},dp\pi}}{4\sqrt{2}}  \\
                 0 & 
                 \frac{\left(t_{\text{Se},dp\sigma}-2 \sqrt{3} t_{\text{Se},dp\pi}\right)}{4\sqrt{2}}  & 
                 \frac{t_{\text{Se},dp\sigma} + 2 \sqrt{3} t_{\text{Se},dp\pi}}{4\sqrt{2}}  \\
                 0 & \frac{\sqrt{3} t_{\text{Se},dp\sigma}}{2\sqrt{2}}  & \frac{\sqrt{3} t_{\text{Se},dp\sigma}}{2\sqrt{2}}  \\
                 \frac{t_{\text{Se},dp\pi}}{\sqrt{2}} & 0 & 0 \\
                 \frac{t_{\text{Se},dp\pi}}{\sqrt{2}} & 0 & 0 \\
            \end{array}
            \endgroup
        \right).
    \end{array}
\end{equation}
The rest of the metal-to-ligand matrices needed can be generated in the same way, but note that they are also either a $C_2$ or $C_4$ rotation away from the ones above.

\section{S3: Additional DFT results} \phantomsection \label{appx:additional_dft_res}

\subsection{Spin split bands and AM DOS}

\begin{figure}
    \centering
    \includegraphics[width=\linewidth]{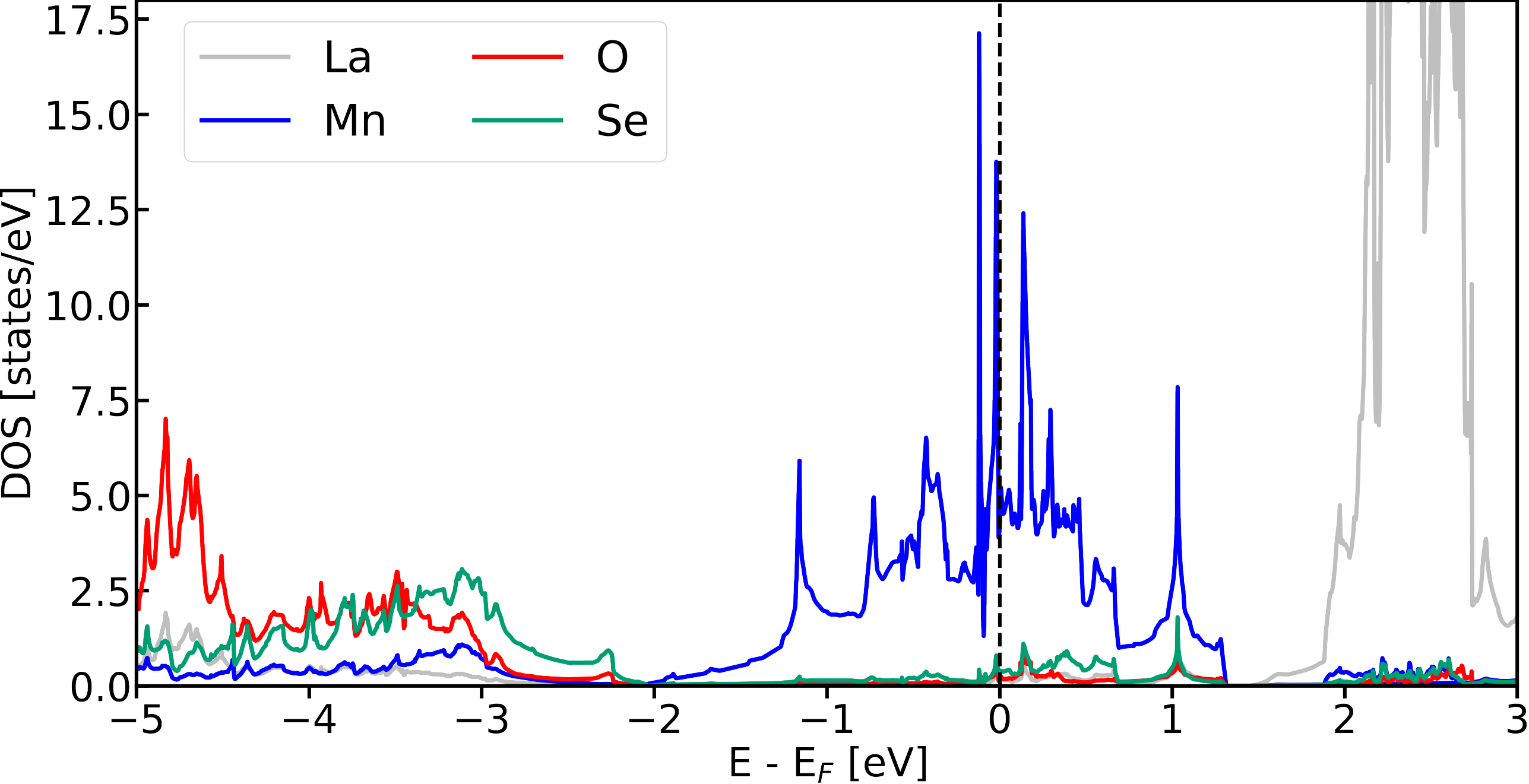}
    \caption{Atomic species resolved density of states for \ce{La2O3Mn2Se2} obtained within a GGA calculation via FPLO. The majority of the states that are located around the Fermi energy belong to the \ce{Mn}, whereas the \ce{O} and \ce{Se} are occupied and entangled.}
    \label{fig:PDOS}
\end{figure}

The species resolved DOS, for non-magnetic $\mathrm{GGA}$ run, without $U$, is shown in Fig.~\ref{fig:PDOS}. We see the dominant Mn character around the fermi level. The Mn states hybridize both with O and Se, which are mainly peaked lower in energy. The dominant Mn states are half-field and they gap out when $U$ is included.

\begin{figure}
    \centering  \includegraphics[width=\linewidth]{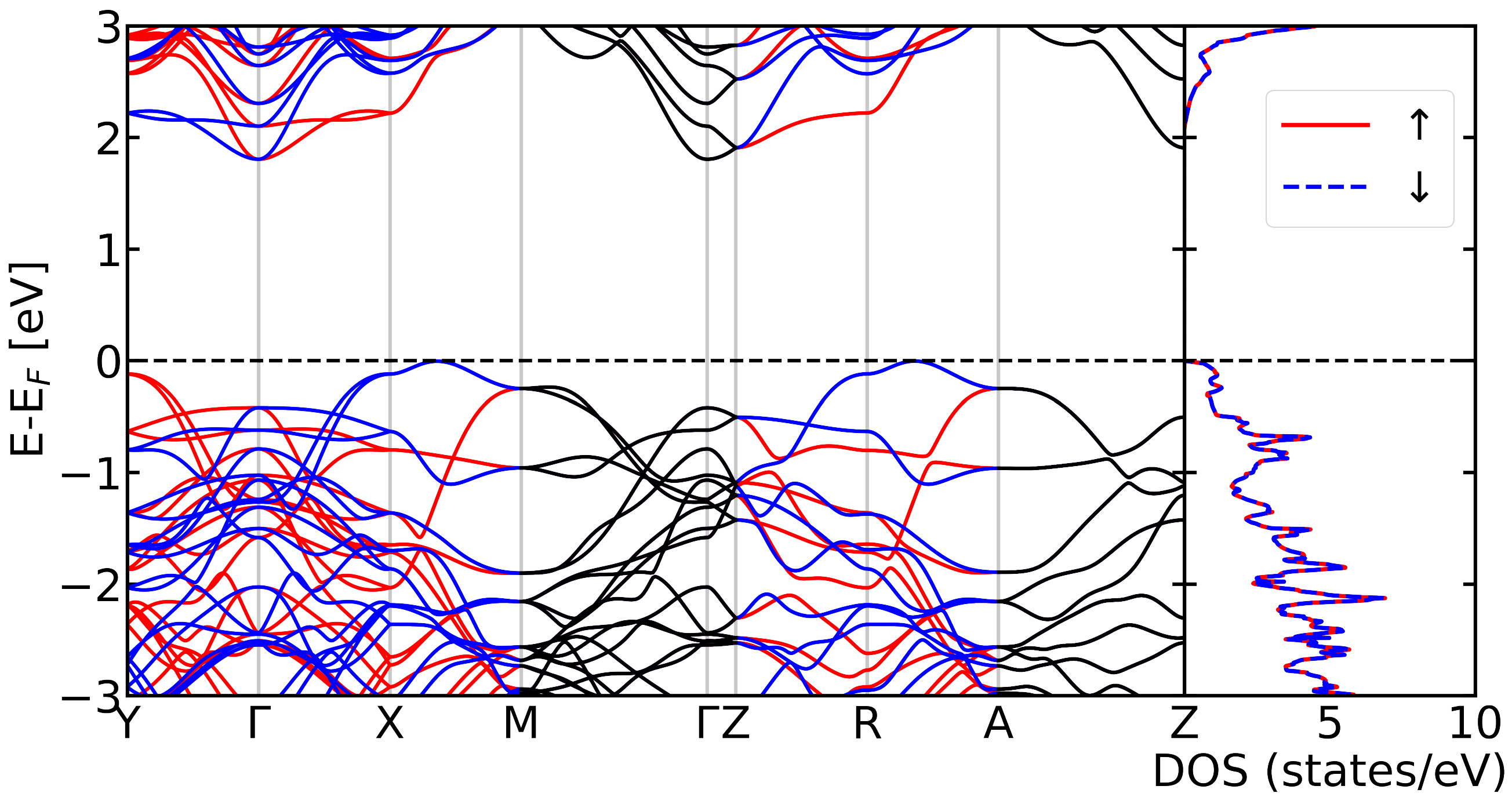}
    \caption{Spin resolved GGA+U band structure and spin polarized density of states of \ce{La2O3Mn2Se2}, and with \(U = 5 \text{eV}\), \(J_H = 1 \text{eV}\) on the \ce{Mn} atoms. Along certain high-symmetry lines, the bands are split (red and blue for spin up and down respectively) and one can observe its inversion behavior (e.g. \textbf{Y}-\boldsymbol{$\Gamma$}-\textbf{X}), whereas along other high-symmetry lines, these are spin-degenerate (black) (e.g. \textbf{M}-\boldsymbol{$\Gamma$}). On the right, the density of states shows the AFM behavior with equal spin up and spin down populations.}
    \label{fig:bands_dos}
\end{figure}

The bands and spin resolved DOS are shown in Fig.~\ref{fig:bands_dos} for the G-AFM state in GGA+U. The DOS have overlapping up and down spin channels, characteristic of an AFM ground state. The AM shows up in the spin resolutions of the bands, showing spin split $k$-paths along certain high-symmetry lines. Due to the position of the \ce{Mn} atoms in the unit cell- one at \((\frac{1}{2},0,0)\) and the other at \((0,\frac{1}{2},0)\)- the spin splitting is expected to be maximized along the main axes and to vanish along the diagonals. Examining the $\mathbf{Y}$-$\bm{\Gamma}$-$\mathbf{X}$ path, we see a spin inversion between $\mathbf{Y}$-$\bm{\Gamma}$ and $\bm{\Gamma}$-$\mathbf{X}$. Along $\mathbf{M}$-$\bm{\Gamma}$ we see the split vanish, because parallel to this high symmetry line is where the two vertical nodal planes lie, \(k_x = \pm k_y\). 

%In the spin resolved DOS the up and down spin channels have the same value, characteristic of an AFM ground state. However, the spin weight is not equally distributed in the bands, with the bands being spin split along certain high-symmetry lines. Due to the position of the \ce{Mn} atoms in the unit cell- one at \((\frac{1}{2},0,0)\) and the other at \((0,\frac{1}{2},0)\)- the spin splitting is expected to be maximized along the main axes and to vanish along the diagonals. This is exactly what we observe in Fig. \ref{fig:bands_dos}, by examining the \textbf{Y}-\boldsymbol{$\Gamma$}-\textbf{X} path, we see a spin inversion between \textbf{Y}-\boldsymbol{$\Gamma$} and \boldsymbol{$\Gamma$}-\textbf{X}, which is the hallmark of AM and, from \textbf{M-\boldsymbol{$\Gamma$}}, we see the split vanish. Parallel to this high symmetry line is where the two vertical nodal planes lie, \(k_x = \pm k_y\). The results do not include SOC, since altermagnetism is not a relativistic effect and therefore we do not require it for AM to emerge.

\subsection{Hopping matrices from DFT} \label{app:eff_direct_hopp_matrices}
In the main text the numerical NN hoppings are shown. Here we present the numerical hopping matrices for other bonds needed and discussed.

\subsubsection{Metal to metal NNN hopping}
For NNN$_a$ bonds, formed on $\mathrm{Mn}_1 - \mathrm{Mn}'_1$, the effective hopping matrix is
\begin{equation} \label{eq:t_B2_effective}
    \begin{array}{l}
        \widetilde{\mathbb{T}}_{\text{NNN}_{a}}= \\[6pt]
        \begin{matrix}
            \begingroup
                \arraycolsep=2pt
                \begin{array}{rrrrr}
                    \makebox[\widthof{$-161.0$}][r]{${x^2\!\!-\!y^2}$} &
                    \makebox[\widthof{$-59.6$}][r]{${z^2}$} & 
                    \makebox[\widthof{$23.5$}][r]{${yz}$} & 
                    \makebox[\widthof{$-194.5$}][r]{${zx}$} & 
                    \makebox[\widthof{$-161.4$}][r]{${xy}$}
                \end{array} 
            \endgroup &
            \begin{matrix} 
                \makebox[\widthof{${x^2\!\!-\!y^2}$}]{\phantom{0}} 
            \end{matrix} \\[4pt]
            \left(
            \begingroup
                \arraycolsep=2pt
                \begin{array}{rrrrr}
                    -161.0 &  77.6 &\mzv  &  \mzv  &  \mzv\\
                     77.6  & -59.6 &\mzv  &  \mzv  &  \mzv\\
                    \mzv   & \mzv  & 23.5 &  \mzv  &  \mzv\\
                    \mzv   & \mzv  &\mzv  & -194.5 &  \mzv\\
                    \mzv   & \mzv  &\mzv  &  \mzv  & -161.4
                \end{array}
            \endgroup
            \right) &
            \begingroup
                \arraycolsep=-20pt
                \begin{array}{l} 
                    {x^2\!\!-\!y^2} \\ {z^2} \\ {yz} \\ {zx} \\ {xy} 
                \end{array}
            \endgroup
        \end{matrix}
    \end{array}
\end{equation}
in units of meV, and the direct is
\begin{equation} \label{eq:t_B2a}
    \begin{array}{l}
        \mathbb{T}_{\text{NNN}_{a}} = \\[6pt]
        \begin{matrix}
            \begin{array}{rrrrr}
                \makebox[\widthof{${x^2\!\!-\!y^2}$}][r]{${x^2\!\!-\!y^2}$}  &
                \makebox[\widthof{$-68.9$}][r]{${z^2}$} & 
                \makebox[\widthof{$1.7$}][r]{${yz}$} & 
                \makebox[\widthof{$-48.4$}][r]{${zx}$} & 
                \makebox[\widthof{$-46.3$}][r]{${xy}$}
            \end{array} &
            \begin{matrix} 
                \makebox[\widthof{${x^2\!\!-\!y^2}$}]{\phantom{0}} 
            \end{matrix} \\[4pt]
            \left(
            \begin{array}{rrrrr}
                \makebox[\widthof{${x^2\!\!-\!y^2}$}][r]{$103.4$}  
                      & -68.9  &\mzv & \mzv  & \mzv\\
                -68.9 &   24.4 &\mzv & \mzv  & \mzv\\
                \mzv  &  \mzv  & 1.7 & \mzv  & \mzv\\
                \mzv  &  \mzv  &\mzv & -48.4 & \mzv\\
                \mzv  &  \mzv  &\mzv & \mzv  & -46.3
            \end{array}\right) &
            \begingroup
                \arraycolsep=-18pt
                \begin{array}{l} 
                    {x^2\!\!-\!y^2} \\ {z^2} \\ {yz} \\ {zx} \\ {xy} 
                \end{array}
            \endgroup
        \end{matrix}
    \end{array}
\end{equation}
in units of meV. For the NNN$_b$ bonds, formed on $\mathrm{Mn}_1 - \mathrm{Mn}''_1$, the effective and direct hopping matrix results are
\begin{equation} \label{eq:t_B3_effective}
    \begin{array}{l}
        \widetilde{\mathbb{T}}_{\text{NNN}_{b}}= \\[6pt]
        \begin{matrix}
            \begin{array}{rrrrr}
                \makebox[\widthof{${x^2\!\!-\!y^2}$}][r]{${x^2\!\!-\!y^2}$} &
                \makebox[\widthof{$-190.3$}][r]{${z^2}$} & 
                \makebox[\widthof{$-11.9$}][r]{${yz}$} & 
                \makebox[\widthof{$39.0$}][r]{${zx}$} & 
                \makebox[\widthof{$9.3$}][r]{${xy}$}
            \end{array} &
            \begin{matrix} 
                \makebox[\widthof{${x^2\!\!-\!y^2}$}]{\phantom{0}} 
            \end{matrix} \\[4pt]
             \left(
            \begin{array}{rrrrr}
                \makebox[\widthof{${x^2\!\!-\!y^2}$}][r]{$144.8$}   
                      &   65.1 & \mzv  &\mzv  &\mzv\\
                 65.1 & -190.3 & \mzv  &\mzv  &\mzv\\
                \mzv  &  \mzv  & -11.9 &\mzv  &\mzv\\
                \mzv  &  \mzv  & \mzv  & 39.2 &\mzv \\
                \mzv  &  \mzv  & \mzv  &\mzv  & 9.3
            \end{array}
            \right) &
            \begingroup
                \arraycolsep=-18pt
                \begin{array}{l} 
                    {x^2\!\!-\!y^2} \\ {z^2} \\ {yz} \\ {zx} \\ {xy} 
                \end{array}
            \endgroup
        \end{matrix}
    \end{array}
\end{equation}
\begin{equation} \label{eq:t_B2b}
    \begin{array}{l}
        \mathbb{T}_{\text{NNN}_{b}}= \\[6pt]
        \begin{matrix}
            \begin{array}{rrrrr}
                \makebox[\widthof{${x^2\!\!-\!y^2}$}][r]{${x^2\!\!-\!y^2}$}  &
                \makebox[\widthof{$-116.7$}][r]{${z^2}$} & 
                \makebox[\widthof{$-75.2$}][r]{${yz}$} & 
                \makebox[\widthof{$20.5$}][r]{${zx}$} & 
                \makebox[\widthof{$44.0$}][r]{${xy}$}
            \end{array} & 
            \begin{matrix} 
                \makebox[\widthof{${x^2\!\!-\!y^2}$}]{\phantom{0}} 
            \end{matrix} \\[4pt]
            \left(
            \begin{array}{rrrrr}
                \makebox[\widthof{${x^2\!\!-\!y^2}$}][r]{$27.0$}
                      &  -29.3 & \mzv  &\mzv  &\mzv\\
                -29.3 & -116.7 & \mzv  &\mzv  &\mzv\\
                \mzv  &  \mzv  & -75.2 &\mzv  &\mzv\\
                \mzv  &  \mzv  & \mzv  & 20.5 &\mzv\\
                \mzv  &  \mzv  & \mzv  &\mzv  & 44.0
            \end{array}
            \right) &
            \begingroup
                \arraycolsep=-18pt
                \begin{array}{l} 
                    {x^2\!\!-\!y^2} \\ {z^2} \\ {yz} \\ {zx} \\ {xy} 
                \end{array}
            \endgroup
        \end{matrix}
    \end{array}
\end{equation}
in units of meV. 

\subsubsection{Metal to ligand hopping}
For Mn to O bond we have
\begin{equation} \label{eq:t_MnO}
    \mathbb{T}_{\text{Mn}_1\shyph\text{O}} \!=\! 
    \begin{matrix}
        \begin{array}{rrr}
            \makebox[\widthof{$1124.2$}][r]{${p_x}$} &
            \makebox[\widthof{$-925.08$}][r]{${p_y}$} & 
            \makebox[\widthof{$-847.0$}][r]{${p_z}$} 
        \end{array} &
        \begin{matrix} 
            \makebox[\widthof{${x^2\!\!-\!y^2}$}]{\phantom{0}} 
        \end{matrix} \\[4pt]
        \left(
        \begin{array}{rrr}
            1124.2 &  \mzv   &  \mzv  \\
            -645.5 &  \mzv   &  \mzv  \\
             \mzv  &  \mzv   &  \mzv  \\
             \mzv  &  \mzv   & -847.0 \\
             \mzv  & -925.0  &  \mzv
        \end{array}
        \right) &
        \begingroup
            \arraycolsep=-18pt
            \begin{array}{l} 
                {x^2\!\!-\!y^2} \\ {z^2} \\ {yz} \\ {zx} \\ {xy} 
            \end{array}
        \endgroup
    \end{matrix}
\end{equation}
in units of meV, and for Mn to Se bond we have
\begin{equation} \label{eq:t_MnSe}
    \mathbb{T}_{\text{Mn}_1\shyph\text{Se}} \!=\! 
    \begin{matrix}
        \begin{array}{rrr}
            \makebox[\widthof{$-293.6$}][r]{${p_x}$} &
            \makebox[\widthof{$-275.9$}][r]{${p_y}$} & 
            \makebox[\widthof{$-504.5$}][r]{${p_z}$} 
        \end{array} &
        \begin{matrix} 
            \makebox[\widthof{${x^2\!\!-\!y^2}$}]{\phantom{0}} 
        \end{matrix} \\[4pt]
        \left(
        \begin{array}{rrr}
             \mzv  &  -275.9 & -504.5 \\
             \mzv  &   421.3 & -106.7 \\
             \mzv  &   511.1 &  462.0 \\
            -233.5 &   \mzv  &  \mzv  \\
            -293.6 &   \mzv  &  \mzv 
        \end{array}
        \right) &
        \begingroup
            \arraycolsep=-18pt
            \begin{array}{l} 
                {x^2\!\!-\!y^2} \\ {z^2} \\ {yz} \\ {zx} \\ {xy} 
            \end{array}
        \endgroup
    \end{matrix}
\end{equation}
in units of meV

\section{S4: Long range exchange model}\label{appx:extended_spin_model}

\begin{figure}
    \centering
    \includegraphics[width=\linewidth]{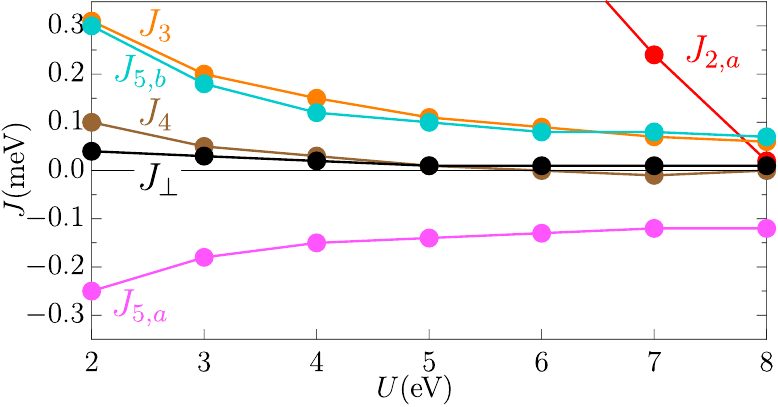}
    \caption{Obtained isotropic exchange parameters via TEMA, for the extended Heisenberg model, as a function of Hubbard $U$. Hund's coupling $J_{H}$ is fixed at 1 eV. The long range exchanges remain small and are plotted in a zoomed in window. See main text Fig.~\ref{fig:TEMA_J123} for the much larger NN and NNN$_{a(b)}$ exchanges.}
    \label{fig:TEMA_extended}
\end{figure}

In order to ensure that the effective Heisenberg Hamiltonian including only the first three nearest exchange parameters $J_1$, $J_{2,a}$, and $J_{2,b}$ describes the system sufficiently well, we extend the Hamiltonian up to the 6$^{\text{th}}$ NN bonds in the $\uvec{a}_1\shyph\uvec{a}_2$ plane including the shortest perpendicular bond, i.e., along the crystallographic $\uvec{a}_3$-direction.
The perpendicular bond is included to resolved the magnetic order in this direction which is seen by neutron experiments~\cite{Cava}.
All bonds included in the TEMA calculation  are listed in Table~\ref{tab:bond_extended}.
Interestingly, the 6NN bond splits similarly to the $J_2$ bonds into two symmetry non-equivalent contributions. 
This results in a Heisenberg Hamiltonian with eight exchange couplings.
This check is mandatory since unresolved longer range contributions are effectively averaged into the $J_1$, $J_{2,a}$, and $J_{2,b}$ fitted parameters by the nature of least squares fit. 
The results as a function of Hubbard $U$ on the strongly localized Mn $d$-orbitals are summarized in Fig.~\ref{fig:TEMA_extended}.
The estimate shows that the longer range contributions are very small and can be neglected in the effective spin Hamiltonian.

\begin{table}[t!]
    \centering
    \begin{tabular}{c c c}
    \hline
    \hline
    Label & Distance [\AA]  & Bond type \\
    \hline
     $J_1$ & 2.9185 & Mn-Mn\\
     $J_{2,a}$ & 4.1274 & Mn-O-Mn\\
     $J_{2,b}$ & 4.1274 & Mn-Se-Mn\\
     $J_3$ & 5.8370 & Mn-Mn \\
     $J_4$ & 6.5260 & Mn-Mn\\
     $J_{5,a}$ & 8.2548 & Mn-O-Mn-O-Mn\\
     $J_{5,b}$ & 8.2548 & Mn-Se-Mn-Se-Mn\\
     $J_{\perp}$ & 9.4026 & Mn-Mn (intraplane)\\
    \hline
    \hline
    \end{tabular}
    \caption{Geometry and assignment of the extended Heisenberg model. Here, we only distinguish the bond type in case there are two symmetry non-equivalent bond types with the same distance.}\label{tab:bond_extended}
\end{table}

\section{S6: Linear spin wave theory}\label{appx:lswt}

We calculate the magnon bands in linear spin wave theory (LSWT). In LSWT we start first from he classical state of the spin model, written in the global $\mathbf{xyz}$ basis, the same basis the spin Hamiltonian has been written in. The classical state has a minimal magnetic unit cell. For each sublattice $\mu$ of the magnetic unit cell we first rotate from the global $\mathbf{xyz}$ basis to a local, site dependent, $\tilde{\mathbf{x}}\tilde{\mathbf{y}}\tilde{\mathbf{z}}_{\mu}$ basis, such that $\tilde{\mathbf{z}}_{\mu}$ is aligned with $\mathbf{S}_{\mu,i}$ on the $\mu^{\text{th}}$ sublattice of the $i^{\text{th}}$ magnetic unit cell. 

In the local basis, the classical state is a fully saturated state along $\tilde{\mathbf{z}}_{\mu}$. To incorporate leading quantum mechanical effects we expand the classical state, writen in the local basis, into boson modes, that describe the fluctuation away from the fully saturated $\tilde{\mathbf{z}}_{\mu}$ state. This is achieved by the Holstein-Primakoff transformation
\begin{equation}
\begin{array}{c}
    \tilde{S}^{z}_{\mu,i}=S-a_{\mu,i}^\dagger a_{\mu,i}, \\[4pt]
    \tilde{S}^{+}_{\mu,i}=\sqrt{2S}\sqrt{1-\dfrac{a_{\mu,i}^\dagger a_{\mu,i}}{2S}}a_{\mu,i} \simeq \sqrt{2S} a_{\mu,i}, \\[8pt]
    \tilde{S}^{-}_{\mu,i}=\sqrt{2S}a_{\mu,i}^\dagger \sqrt{1-\dfrac{a_{\mu,i}^\dagger a_{\mu,i}}{2S}} \simeq \sqrt{2S} a_{\mu,i}^\dagger.
\end{array}
\end{equation}
The "linear" in LSWT comes from he approximation of the square root to 1, leaving linear in $S$ contributions only. The bosonic modes $a_i$ are describing the reduction away from the fully saturated $S$ moment in the local $\tilde{\mathbf{z}}_{\mu}$ frame.

Using the linearized Holstein-Primakoff transformation, we can rewrite the original spin Hamiltonian, written in the local basis now, into a bosonic BdG Hamiltonian. Fourier transforming the bosonic operators $a_{\mu,i}$, we have finally 
\begin{equation}
\begin{array}{cc}
    \mathcal{H} = S(S+1)E_{CL} + S \mathcal{H}_{\mathrm{BdG}}, \\[2pt]
\mathcal{H}_{\mathrm{BdG}} = \sum\limits_{\mathbf{k}}\mathcal{A}_{\mathbf{k}}^\dagger H_{\mathbf{k}} \mathcal{A}_{\mathbf{k}}, \\[2pt]
\mathcal{A}_{\mathbf{k}}^\dagger = (a_{\mu,\mathbf{k}}^\dagger a_{\mu,-\mathbf{k}}) 
\end{array}
\end{equation}
where $E_{CL}$ the energy of the classical state. Assuming $N_m$ sublattices in the magnetic unit cell, the BdG Hamiltonain dimensions are $2N_m\times2N_m$.

\subsection{BdG Hamiltonian matrix}
The bosonic BdG Hamiltonians need to fulfill a specific form, which reflects the bosonic nature of the enlarged basis representation resulting in the $2N_m\times2N_m$ dimensional $H_{\mathbf{k}}$ matrix, specifically
\begin{equation}
    H_{\mathbf{k}} = 
    \left(
\begin{array}{cc}
    h_{\mathbf{k}} & \Delta_{\mathbf{k}} \\
    \Delta_{\mathbf{k}}^\dagger & h_{-\mathbf{k}}^*
\end{array}
    \right)
\end{equation}
where the submatrices $h_{\mathbf{k}}$ and $\Delta_{\mathbf{k}}$ are $N_m\times N_m$ dimentional, and have the additional properties $h_{\mathbf{k}} = h_{\mathbf{k}}^\dagger$, and $\Delta_{\mathbf{k}} = \Delta_{-\mathbf{k}}^T$. This form of $H_{\mathbf{k}}$  is such that the BdG Hamiltonian will respect particle-hole symmetry. Due to the enlarged representation, redundantly involving both creation and annihilation operators, there is a built in particle-hole symmetry, which in momentum space amounts to the statement that any bosonic BdG Hamiltonian in momentum space will satisfy the relation $\left( H_{-\mathbf{k}} \right)^* =  \Sigma^x H_{\mathbf{k}} \Sigma^x$, where $\Sigma^x=\sigma^x\otimes\mathbb{1}_{N_m\times N_m}$. This property will also relate the diagonalizing matrix between $\mathbf{k}$ and $-\mathbf{k}$, which can be useful in numerical evaluation of expectation values, as we will see in the following subsection.

It is a strait forward calculation to start from our $J_1$, $J_{2,a}$, $J_{2,b}$ model, with the Neel like state as the classical ground state, and perform the transformations outlined in the previous subsection, to arrive finally at the magnon BdG Hamiltonian with
\begin{equation}
\begin{array}{l}
 h_{\mathbf{k}} \!=\! \left[ 4J_1 \!+\! (J_{2,a} \!+\! J_{2,b})(\cos\mathbf{a}_1\!\!\cdot\!\mathbf{k}\!+\!\cos\mathbf{a}_2\!\cdot\!\mathbf{k}\!-\!2) \right]\sigma^o  \\[6pt]
 \multicolumn{1}{r}{+\left[(J_{2,a} \!-\! J_{2,b})(\cos\mathbf{a}_1\!\!\cdot\!\mathbf{k}\!+\!\cos\mathbf{a}_2\!\cdot\!\mathbf{k}) \right]\sigma^z}, \\[12pt]
  \Delta_{\mathbf{k}} \!=\!  J_1 (1 + \cos\mathbf{a}_1\!\!\cdot\!\mathbf{k}\!+\!\cos(\mathbf{a}_1\!\!+\!\mathbf{a}_2)\!\cdot\!\mathbf{k}\!+\!\cos\mathbf{a}_2\!\cdot\!\mathbf{k})\sigma^x  \\[6pt]
 \multicolumn{1}{r}{+J_1 (\sin\mathbf{a}_1\!\!\cdot\!\mathbf{k}\!+\!\sin(\mathbf{a}_1\!\!+\!\mathbf{a}_2)\!\cdot\!\mathbf{k}\!+\!\sin\mathbf{a}_2\!\cdot\!\mathbf{k})\sigma^y}. 
 \end{array}
\end{equation}

\subsection{Diagonalization of bosonic BdG Hamiltonian}

To find the magnon modes $\gamma_{\mu,\mathbf{k}}$ we need to diagonalize $\mathcal{H}_{\mathrm{BdG}}$. Given that this is a bosonic Hamiltonian, we are looking for a diagonalization method such that the bosonic commutation relations will be preserved, and the eigen states $\gamma_{\mu,\mathbf{k}}$, i.e., the magnons, will also be bosons. This amounts to looking for matrix $J_{\mathbf{k}}$ such that the quantum metric $\Sigma^z=\sigma^z\otimes\mathbb{1}_{N_m\times N_m}$ is preserved, i.e., $J_{\mathbf{k}} \Sigma^z J^\dagger_{\mathbf{k}} = J^\dagger_{\mathbf{k}} \Sigma^z J_{\mathbf{k}} = \Sigma^z$. This is called a paraunitary transformation. 

It can be shown that a paraunitary transformation that diagonalizes $H_{\mathbf{k}}$ is always constructable when $H_{\mathbf{k}}$ is positive definite. Singular isolated momentum points with a positive semidefinate matrix, like the Golston mode around the ordering wave vector, can also be treated. Following section 3 in Ref.~\cite{ColpaPA1978} one can constructively prove the existence of a new basis $\Gamma_{\mathbf{k}}^\dagger = ( \gamma_{\mu,\mathbf{k}}^\dagger\  \gamma_{\mu,-\mathbf{k}})$  related to the original basis by a paraunitary transformation $\Gamma_{\mathbf{k}} = J_{\mathbf{k}} \mathcal{A}_{\mathbf{k}}$ such that
\begin{equation}
\begin{array}{c}
     \mathcal{A}_{\mathbf{k}}^\dagger H_{\mathbf{k}} \mathcal{A}_{\mathbf{k}} = \Gamma_{\mathbf{k}}^\dagger \Omega_{\mathbf{k}} \Gamma_{\mathbf{k}},  \\[4pt] \Omega_{\mathbf{k}} = \mathrm{diag}(\omega_{\mu,\mathbf{k}}, \omega_{\mu,-\mathbf{k}})
\end{array}
\end{equation}
where $\omega_{\mu,\mathbf{k}}$ are one magnon energy bands, corresponding to $\gamma_{\mu,\mathbf{k}}$ bosonic eigenmodes. The constructive proof also gives the algorithm to find such a paraunitary diagonalizing matrix.

\subsection{Chiral magnons from evaluating \texorpdfstring{\(S^{\text{z}}_{\text{tot}}\)}{}}

Having diagonalized the Hamiltonian we can now calculate some quantities of interest. One such quantity is the total spin moment along the $\mathbf{z}$ axis $S^z_{\text{tot}}$. For chiral magnons we will see that one magnon modes split into well defined values of $S^z_{\text{tot}}$, which is a characteristic scenario in the AM.

Since Heisenberg Hamiltonians commute with $S^z_{\text{tot}}$ operator, that means that a single eigen state, i.e. a non-degenerate one magnon state, must have definite values of $S^z_{\text{tot}}$. The $S^z_{\text{tot}}$ operator in momentum basis will read
\begin{equation}
\begin{array}{c}
    S^z_{\text{tot}} = \sum\limits_{\mu,i} S^z_{\mu,i} = \sum\limits_{i}(-1)^{1+\mu}\tilde{S}^z_{\mu,i} \\
    = \sum\limits_{i} (-1)^{\mu}a^\dagger_{\mu,i}a_{\mu,i} 
    = \sum\limits_{\mathbf{k}} (-1)^{\mu}a^\dagger_{\mu,\mathbf{k}}a_{\mu,\mathbf{k}}
\end{array}
\end{equation}
We now wish to switch into the magnon basis so we can easily meassure the operator on the one magnon states. Using the paraunitary relation between $\Gamma_{\mathbf{k}}$ and $\mathcal{A}_{\mathbf{k}}$ 
\begin{equation}
    a_{\mu,\mathbf{k}} = \left[J^{-1}_{\mathbf{k}}\right]^{\mu,\nu} \gamma_{\nu,\mathbf{k}}+ \left[J^{-1}_{\mathbf{k}}\right]^{\mu,N_m+\nu}\gamma^\dagger_{\nu,-\mathbf{k}},
\end{equation}
we can rewrite the expectation value $\langle S^z_{\text{tot}} \rangle$ in the magnon basis
\begin{equation}
\begin{array}{c}
    \langle S^z_{\text{tot}} \rangle =\sum\limits_{\mathbf{k}} (-1)^{\mu}\Big\lbrace
    \left(\left[J^{-1}_{\mathbf{k}}\right]^{\mu,\nu}\right)^*
    \left[J^{-1}_{\mathbf{k}}\right]^{\mu,\nu}
    n^{\gamma}_{\nu,\mathbf{k}} \\
+    \left(\left[J^{-1}_{\mathbf{k}}\right]^{\mu,N_m+\nu}\right)^*
    \left[J^{-1}_{\mathbf{k}}\right]^{\mu,N_m+\nu}
    (1+n^{\gamma}_{\nu,-\mathbf{k}})  
    \Big\rbrace
\end{array}
\end{equation}
where $n^{\gamma}_{\nu,\mathbf{k}} = \langle\gamma^\dagger_{\mu,\mathbf{k}}\gamma_{\mu,\mathbf{k}}\rangle$. Given the one magnon state $\left|\psi_{\lambda,\mathbf{p}}\right\rangle = \gamma_{\lambda,\mathbf{p}}^\dagger \left|0\right\rangle$ we find that the operator expectation value can be calculated as
\begin{equation}
\begin{array}{c}
    \left\langle\psi_{\lambda,\mathbf{p}}\right| S^z_{\text{tot}}\left|\psi_{\lambda,\mathbf{p}}\right\rangle = 
    (-1)^{\mu}\Big\lbrace
    \left(\left[J^{-1}_{\mathbf{p}}\right]^{\mu,\lambda}\right)^*\left[J^{-1}_{\mathbf{p}}\right]^{\mu,\lambda}
    \\
+    \left(\left[J^{-1}_{-\mathbf{p}}\right]^{\mu,N_m+\lambda}\right)^*
    \left[J^{-1}_{-\mathbf{p}}\right]^{\mu,N_m+\lambda}
    \Big\rbrace,
\end{array}
\end{equation}
where we note how both the $\mathbf{k}=\mathbf{p}$ and the $\mathbf{k}=-\mathbf{p}$ momentum give a contribution to the expectation value. We also comment that we have omitted an overall constant (same for any one magnon state) in the above expectation value, which is precisely the value of $S^z_{\text{tot}}$ for the ground state, i.e., the vacuum with no magnons present, which would be 0 in this case. Finally, note that due to the particle-hole symmetry present in any BdG Hamiltonian the opposite momentum paraunitary diagonalization matrices are related by
   $\left( J_{-\mathbf{k}} \right)^* =  \Sigma^x J_{\mathbf{k}} \Sigma^x$. Use of this relation can be important to remove arbitrary diagonalization phases from different momentum points in the above equation when evaluated numerically.